\documentclass[twoside,twocolumn]{article}
\usepackage[super,sort&compress,comma]{natbib}
\usepackage[version=3]{mhchem} 
\usepackage{times}
\usepackage{sectsty}
\usepackage{balance}
\usepackage{graphicx} 
\usepackage{lastpage}
\usepackage[format=plain,justification=raggedright,
singlelinecheck=false,font=small,labelfont=bf,
labelsep=space]{caption}
\usepackage{fancyhdr}
\usepackage{color}
\usepackage{bm}
\usepackage[right]{eurosym}
\usepackage{dcolumn}
\usepackage{amsmath,amssymb} 

\newcommand{\unit}[1]{\ensuremath{\,\mathrm{{#1}}}}
\newcommand{\wn}{\ensuremath{\,\mathrm{{cm^{-1}}}}}

\begin{document}

\thispagestyle{plain}
\fancypagestyle{plain}

\renewcommand{\thefootnote}{\fnsymbol{footnote}}
\renewcommand\footnoterule{\vspace*{1pt}%
\hrule width 3.4in height 0.4pt \vspace*{5pt}}

\makeatletter
\renewcommand\@biblabel[1]{#1}
\renewcommand\@makefntext[1]%
{\noindent\makebox[0pt][r]{\@thefnmark\,}#1}
\makeatother
\renewcommand{\figurename}{\small{Fig.}~}
\sectionfont{\large}
\subsectionfont{\normalsize}

\fancyfoot{}
\fancyfoot[RO]{\footnotesize{\sffamily{1--\pageref{LastPage} ~\textbar  \hspace{2pt}\thepage}}}
\fancyfoot[LE]{\footnotesize{\sffamily{\thepage~\textbar\hspace{4.4cm} 1--\pageref{LastPage}}}}
\fancyhead{}
\renewcommand{\headrulewidth}{1pt}
\renewcommand{\footrulewidth}{1pt}
\setlength{\arrayrulewidth}{1pt}
\setlength{\columnsep}{6.5mm}
\setlength\bibsep{1pt}

\twocolumn[
  \begin{@twocolumnfalse}
\noindent\LARGE{\textbf{Homo- and heteronuclear alkali metal trimers formed
 on helium na\-no\-drop\-lets. II.~Femtosecond spectroscopy and spectra assignments}}
\vspace{0.6cm} 

\noindent\large{\textbf{Christian Giese\textit{$^{a}$}, Frank
    Stienkemeier\textit{$^{a}$}, Marcel Mudrich\textit{$^{a}$}, Andreas W. Hauser\textit{$^{b}$} and
Wolfgang E. Ernst\textit{$^{c}$}}}\vspace{0.5cm}

\vspace{0.6cm}

\noindent \normalsize{ Homo- and heteronuclear alkali quartet trimers
  of the type K$_{3-n}$Rb$_n$ ($n=0,1,2,3$) formed on helium
  nanodroplets are probed by one-color femtosecond photoionization
  spectroscopy. The obtained frequencies are assigned to vibrations in
  different electronic states by comparison to high level \emph{ab
    initio} calculations of the involved potentials including
  pronounced Jahn-Teller and spin-orbit couplings. Despite the fact
  that the resulting complex vibronic structure of the heavy alkali
  molecules complicates the comparison of experiment and theory we
  find good agreement for many of the observed lines for all species.}
\vspace{0.5cm}
 \end{@twocolumnfalse}
  ]


\footnotetext{\textit{$^{a}$~Physikalisches Institut, Universit\"at
      Freiburg, 79104 Freiburg, Germany. E-mail:
      marcel.mudrich@physik.uni-freiburg.de,
      christian.giese@physik.uni-freiburg.de}}

\footnotetext{\textit{$^{b}$~Centre for Theoretical Chemistry and Physics,
		The New Zealand Institute for Advanced Study, Massey University, 0745 Auckland,
    New Zealand. E-mail: andreas.w.hauser@gmail.com}}

\footnotetext{\textit{$^{c}$~Institute of Experimental Physics, Graz
    University of Technology, Petersgasse 16, A-8010 Graz, Austria.
    E-mail: wolfgang.ernst@tugraz.at}}

\section{Introduction}
Helium nanodroplet isolation (HENDI) spectroscopy has become an
established technique for producing and probing molecules, clusters,
and chemical reactions at low
temperatures.\cite{Toennies:2004,Stienkemeier:2006,Tiggesbaumker:2007,
  ErnstHB:2011} Dopant species are cooled down to the internal
temperature of the droplet (0.37 K)\cite{Hartmann:1995} and readily
form complexes, weakly bound ones being favored.\cite{Higgins:1998}
Due to their low binding energy, high-spin alkali-metal molecules are
hardly accessible in conventional gas-phase experiments but are
preferentially formed on helium
droplets.\cite{Higgins:1996,Higgins2:1996,Higgins:1998} Alkali atoms,
molecules and clusters in high-spin states have been extensively
studied by cw\cite{Mudrich:2004,Buenermann:2004,Buenermann:2007,nagl:prl08,Bruehl3:2001,Ernst:Ces2006} and time-resolved spectroscopy.\cite{Schulz:2001,Droppelmann:2009,Schulz:2004,Droppelmann:2004,Mudrich:2008,Stienkemeier:2006,Claas:2006,Claas:2007,Mudrich:2009}

Although helium is the least perturbing solvent and helium
nanodroplets are often referred to as a nearly ideal spectroscopic
matrix, electronic spectra of atoms or molecules in or on He droplets
are still strongly broadened.  On the one hand, this is a severe
limitation as to the precision of ro-vibrational spectra as compared
to gas-phase spectroscopy. On the other hand, valuable information
about the interaction of the dopant with the droplet may be obtained
from helium perturbations.\cite{Grebenev:1998,Dick:2001,AuboeckAymar:2010,Auboeck:jpca2007,Gruner:2011,Schlesinger:2010} Significantly improved spectral resolution can be achieved by time-domain
spectroscopy using the femtosecond pump-probe
technique\cite{Claas:2006,Mudrich:2009} or by measuring fine and
hyperfine structure directly in the microwave regime~\cite{Koch:2009}.

In previous experiments, high-spin alkali trimers formed on helium
nanodroplets have been interrogated using laser
spectroscopy\cite{Higgins2:1996,HigginsJCP:2000,Reho:2001} as well as
sophisticated variations thereof in order to separate spectrally
overlapping bands of various trimer species.\cite{auboeck08,nagl:prl08,Nagl_jcp:2008,hauser10jcp} In the case of Na$_3$ and
K$_3$, excitation and emission spectroscopy allowed for vibrational
resolution.\cite{Higgins2:1996,HigginsJCP:2000,Reho:2001,hauser10jcp}
High-spin alkali trimers are particularly interesting model systems
for studying nonadiabatic transitions from weakly bound van der Waals
aggregates to covalently bound metallic
clusters.\cite{shell-model,Hauser:book2011,Theisen:2011} Upon
electronic excitation of high-spin Na$_3$ and K$_3$ formed on helium
droplets into the $2^4$E$^{\prime}$-state, the quartet trimers undergo
intersystem crossing to the doublet manifold followed by dissociation
into singlet dimers and single atoms.\cite{Higgins:1996,Reho:2001}

Alkali trimers in their doublet states were investigated at very high
resolution, which allowed to derive details of Jahn-Teller and pseudo
Jahn-Teller interactions from the splitting of rotational
states.\cite{pseudorotNa3Ernst,Golonzka:2004} As trimers in their
weakly bound quartet states have only been observed on helium
droplets, such high resolution has not been achieved in the optical
spectroscopy range. As a complementary approach to spectroscopy with
vibronic resolution the femtosecond pump-probe technique can be
applied, as was demonstrated for the doublet trimers Na$_3$ and
K$_3$.\cite{Baumert:1993,Ruppe:1996,Reischl:1995,Reischl:1996,
  Rutz:1997,Schreiber} This technique relies on the creation and
probing of a coherent superposition of vibrational states (wave
packet, WP) in the time domain. Subsequent Fourier analysis yields
information about difference frequencies between the coherently
excited states. The advantage with respect to cw spectroscopy is that
spectral information can be gained even in strongly perturbed systems
that are subject to fast dissociation or to interactions with an
environment. Thus, by using femtosecond pump-probe spectroscopy, it
was possible to measure vibrational frequencies of the previously
unobserved dissociative B-state of the K$_3$
trimer,\cite{Ruppe:1996,Schreiber} which was later confirmed in
\emph{ab initio} calculations.\cite{Hauser:jcp2008} At high laser
intensity, additional vibrational WP dynamics in the ground state was
observed induced by resonant impulsive Raman scattering
(RISRS).\cite{Baumert:1993,Rutz:1997,Schreiber} The femtosecond
dynamics of Na$_3$ in the B-state was found to be dominated by the
symmetric stretch mode Q$_s$ in contrast to cw
measurements.~\cite{Reischl:1995,Rutz:1997,Schreiber} This mode
decays within a few picoseconds due to intramolecular vibrational
redistribution (IVR) to the two other modes. Note, however, that the
excitation of vibrational modes was found to be quite selective
depending on the laser pulse parameters, in particular the pulse
duration.\cite{Baumert:1993,Rutz:1997,Ruppe:1996,Reischl:1996} On the
one hand, this offers the opportunity of accessing vibrational modes
that are not excited by cw laser excitation, in particular when using
short and intense pulses.\cite{Baumert:1993,Rutz:1997} On the other
hand, this can hamper the direct comparison of the Fourier spectra
obtained from femtosecond or picosecond pump-probe transients with cw
excitation spectra. Besides, additional fine structure effects induced
by \textit{e.\,g.} spin-orbit coupling, rotations, or matrix
interactions, which naturally appear as line splittings, shifts and
broadenings in absorption or emission spectra, are not directly probed
by pump-probe vibrational coherence
spectroscopy.\cite{Gruebele:1992,Mudrich:2009} Femtosecond pump-probe
spectroscopy has been applied to alkali atoms and molecules attached
to helium nanodroplets by our group in a series of
experiments.\cite{Stienkemeier:2006,Claas:2006,Claas:2007,Mudrich:2009} Besides obtaining a wealth of
dynamical information, high-resolution spectra of diatomics have been
extracted that give insight into the subtle couplings of the vibrating
molecules to the helium droplet
surface.\cite{Schlesinger:2010,Gruner:2011}

Due to the small size and the simple valence structure of the
constituents, alkali trimers are particularly attractive as benchmark
systems for accurate theoretical treatment. Besides the mentioned
spin-orbit couplings the Jahn-Teller (JT) effect, specifically the
problem of E$\otimes$e vibronic coupling, has been a matter of high
interest and continuous research.\cite{hauser10jcp,hauser10cp} The
calculation of accurate potential energy surfaces (PES) for trimers
formed of the heavy alkali elements is computationally quite
demanding.  However, with the advances in computational power and
techniques, PES for homo- and heteronuclear alkali trimers involving
potassium, rubidium and cesium have recently been determined by
several groups.\cite{Quemener-KK2,nagl:prl08,hauser10jcp,hauser10cp,
  Hauser:book2011} Besides, alkali trimers in quartet states are
particularly well suited for studying nonadditive three-body
interaction terms, which represent significant contributions to the
binding energy.\cite{HigginsJCP:2000,Soldan:2003}

Recent advances in forming ultracold molecules by photo- or
magneto-association through Feshbach resonances using trapped samples
of ultracold atoms has stimulated increasing demand for accurate PES
of three-atomic systems. In particular, schemes for stabilizing
initially highly excited molecules into low-lying levels are based on
the precise knowledge of molecular structure and
spectra.\cite{Lang:2008,Ni:2008,Danzl:2010} Moreover, accurate
triatomic PES are required in the emerging field on ultracold
chemistry.\cite{Krems:2008,Bell:2009,Hutson:2010,Knoop:2010,Ospelkaus:2010}

In this work we present one-color femtosecond pump-probe spectroscopy
of homo- and heteronuclear quartet alkali trimers of rubidium and
potassium atoms. Vibrational frequencies of excited and ground state
trimers in all possible combinations of species are extracted and
compared to high-level quantum chemical calculations. In addition, the
real-time dynamics of individual frequency components is analyzed. The
article is organized as follows: After a short description of the
apparatus the experimental results are presented and data analysis is exemplified for Rb$_3$. This is followed by a theoretical part summarizing the findings of our previous \emph{ab initio} calculations
on these systems\cite{hauser10cp,shell-model,hauser10jcp} and makes
extensive use of the data provided in the closely connected
Ref.~\citenum{paper1}, which is referred to as Paper~I throughout this
article. The final section is dedicated to the discussion and interpretation
 for all studied trimers Rb$_3$, KRb$_2$, K$_2$Rb and
K$_3$ are presented based on the theoretical spectra simulations.

\section{Experimental}
\label{sec:experimental}

The details of the experimental setup have been described in previous
publications.\cite{Mudrich:2009,Mudrich:2004} Helium droplets are
formed in the expansion of helium at high pressure ($\sim 50$\,bar)
out of a cold nozzle ($T=15\unit{K}$, diameter $d=5\unit{\mu m}$). In
two stainless steel pickup cells, a number Rb and K atoms are attached
to the droplets according to the pickup statistics at a given vapor
pressure.\cite{Buenermann:2011} Doping-cell temperatures are adjusted
to achieve highest formation rates of trimer species and are listed in
Table~\ref{tab:oventemps}.  \newline Alkali atoms and molecules are
peculiar dopants in that they reside in weakly bound dimple-like
states at the surface of helium nanodroplets.\cite{Dalfovo:1994,
  Ancilotto:1995,Higgins:1996,Stienkemeier:1996} Because of the high
mobility on the helium surface homonuclear clusters (Rb$_3$ and K$_3$)
and heteronuclear clusters (KRb$_2$ and K$_2$Rb) are formed when
doping one droplet with three atoms. Upon cluster formation, the
binding energy is dissipated by evaporation of helium atoms which
leads to the depletion of strongly bound clusters and to an enrichment
of the weakly bound high-spin clusters, \emph{i.\,e.} alkali trimers
in the lowest quartet state. According to recent simulations of the formation
of alkali clusters on helium droplets\cite{Buenermann:2011}, we expect a contribution 
of doublet trimers on the order of $\sim$10\% with respect to the high spin molecules.
However, so far no low spin trimer states have been identified in 
cw spectra of alkali trimers attached to helium droplets. Therefore, all 
observed frequencies discussed in section\,\ref{sec:analysis} are related to quartet states. 
In the presented femtosecond pump-probe measurements, we can
not definitely rule out doublet trimers to contribute to the spectra.\newline
As an example, the doublet ground state 
have 
potential of Rb$_3$ ($1^2$E$^{\prime}$) has a formation energy of
about 5321\wn{} as compared to 939\wn{} for the
$1^{4}$A$_{2}^{\prime}$ lowest quartet
state.\cite{hauser10cp,hauser10jcp} Thus-formed trimers immediately
thermalize to the nanodroplet equilibrium temperature (370\,mK) such
that only the vibrational ground state ($v=0$) and a few rotational
states are populated. Thus, helium nanodroplet isolation (HENDI) is an
ideal method for species and state preparation to perform single
quantum channel experiments.

\begin{table}[ht]
\small
\centering
	\caption{Pick up cell temperatures for optimum trimer formation.}
\begin{tabular*}{0.45\textwidth}{@{\extracolsep{\fill}}lcc} \hline
                  Species&$T_{Rb-oven}$ [$^{\circ}$~C]&$T_{K-oven}$ [$^{\circ}$~C]\\
                  \hline
                   Rb$_{3}$	&115 &-\\
                   Rb$_{2}$K	&103 &115\\
                   RbK$_{2}$	&83  &130\\
                   K$_{3}$	&-   &140\\ \hline	
 \end{tabular*}		
\label{tab:oventemps}
\end{table}

The laser system consists of a Ti:Sapphire femtosecond oscillator
(Chameleon series/Coherent) that produces Fourier limited pulse trains
over a broad spectral range of 700--1050\,\unit{nm} (9500--14300\wn{})
at 80\,\unit{MHz} pulse repetition
rate with a cw output power of 1-3\,W. For the chosen excitation
frequencies, the laser pulse lengths are about 160\,\unit{fs}, yielding
a spectral full width at half maximum (FWHM) of about 80\wn{}. Pulse
pairs are created with a Mach-Zehnder type interferometer where a
commercial delay stage (Newport) is placed in one of the arms. In this
way the time delay between the pulses is controlled with
sub-femtosecond precision. The droplet beam is crossed with the laser
beam at ninety-degree angle inside the active volume of a commercial
quadrupole mass spectrometer (Extrel) and the formed molecular ions
are recorded in counting mode as a function of the delay time.

\subsection{Vibrational femtosecond pump-probe spectroscopy of
  triatomic molecules}
  \label{sec:ppspectra}
  Vibrational WP dynamics is probed by applying pairs of identical
  laser pulses with variable delay.  The first `pump' pulse excites a
  coherent superposition of vibrational states
  in a well defined region of the molecular PES according to the laser
  bandwidth and the overlap of ground and excited state wave functions
  (Franck-Condon (FC)-factors).
  The created WP starts to propagate on the PES until the second
  `probe' pulse projects it onto a detection state (ionic or
  fluorescent) after a delay time $\Delta t$. According to the
  Franck-Condon principle, transitions between electronic states are
  energetically allowed at molecular coordinates at which the laser
  energy matches the energy gap between ground and excited state PES
  (`Condon points') and transition probabilities are determined by the
  FC-factors. Well-localized FC-regions in coordinate space
  (`FC-windows') are essential for filtering out the WP dynamics and
  for obtaining oscillatory ion yields as a function of $\Delta t$ with
  high contrast.\cite{Claas:2006} Such regions are mostly located at
  classical turning points of the molecular vibration.

  Based on previous measurements of vibrational WP dynamics in
  potassium and rubidium dimers, we expect essentially two possible
  excitation pathways to contribute to the pump-probe signal.
  \cite{Claas:2006,Mudrich:2009} In
  the first case, the pump pulse populates states in an excited
  quartet potential, and the probe pulse ionizes the molecule by a
  two-photon transition. In the second case, the high intensities of
  the ultrashort laser pulses induce resonant impulsive stimulated
  Raman scattering (RISRS) which creates WPs in the initial quartet
  ground state. Since the vibrational WP oscillates in the symmetry
  adapted coordinates of the molecules, the recorded ion yield is also
  periodically modulated, depending on the number of Condon points and
  on FC-factors. In a harmonic potential the oscillation is undamped
  and will therefore only be limited by the lifetime of the involved
  vibronic states, whereas in anharmonic PES the WP undergoes
  dispersion. This leads to the spreading of the WP and to the loss of
  contrast of the pump-probe signal.  This process is reversible,
  though, and after characteristic delay times the WP rephases to form
  so-called revivals, as nicely seen in pump-probe spectra of
  Rb$_2$.\cite{Mudrich:2009} Irreversible dephasing can be induced
  either by population decay from the initially prepared vibronic
  states by spontaneous emission, inter-system crossing, internal
  conversion, or by dissipative interactions with the environment
  (the helium droplets in our case). In addition, system-bath
  coupling may cause pure dephasing by elastic interactions without
  dissipation.\cite{Gruner:2011}

\begin{figure}[ht]
\begin{center}
\includegraphics[width=0.45\textwidth]{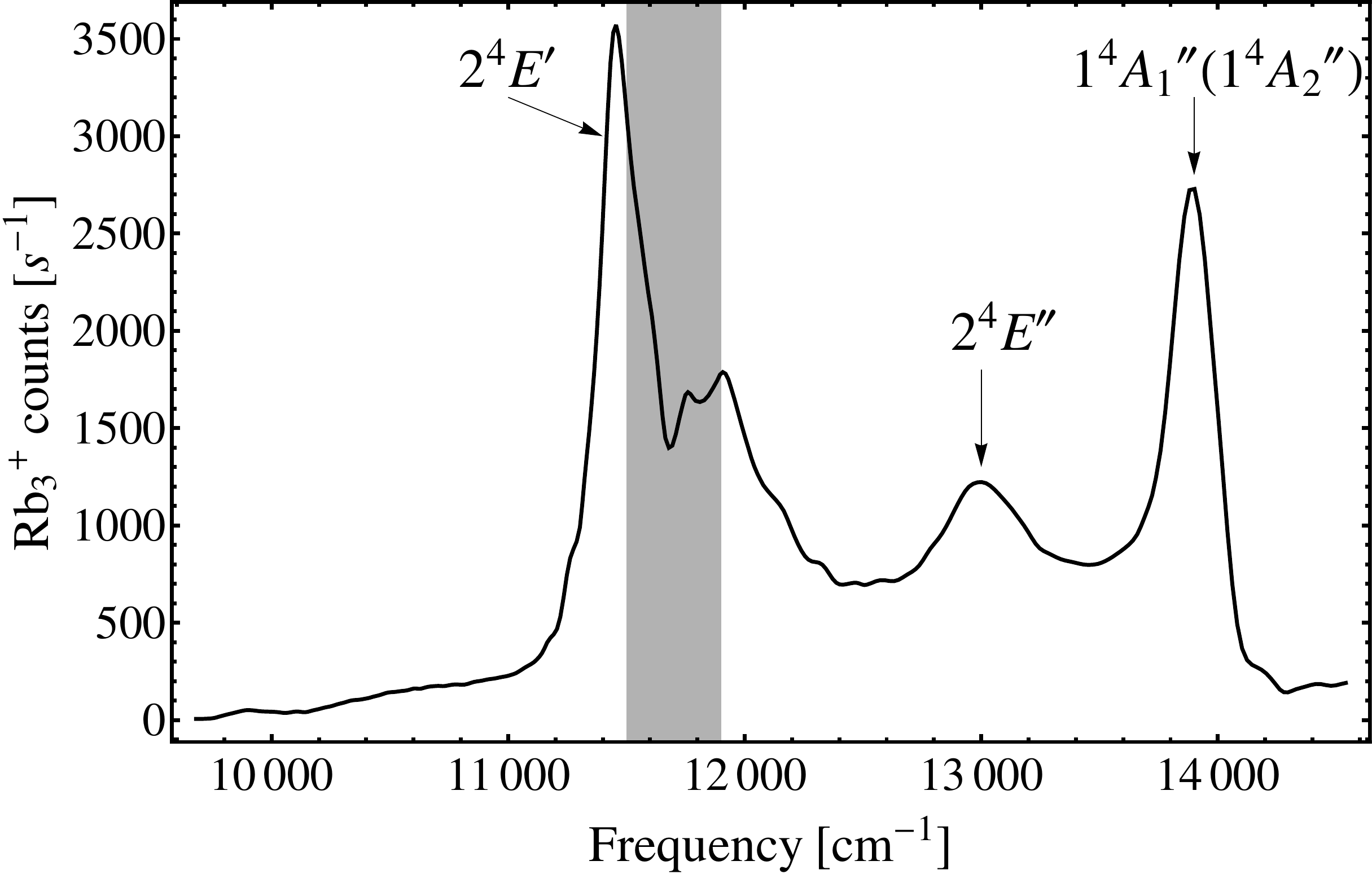}
\caption{Photoionization spectrum of rubidium trimers Rb$_3$ recorded by
  tuning the femtosecond laser. WP oscillations were observed within the shaded area.}
\label{fig:PIRb3}
\end{center}
\end{figure}

\begin{figure}[ht]
\begin{center}
  \includegraphics[width=0.45\textwidth]{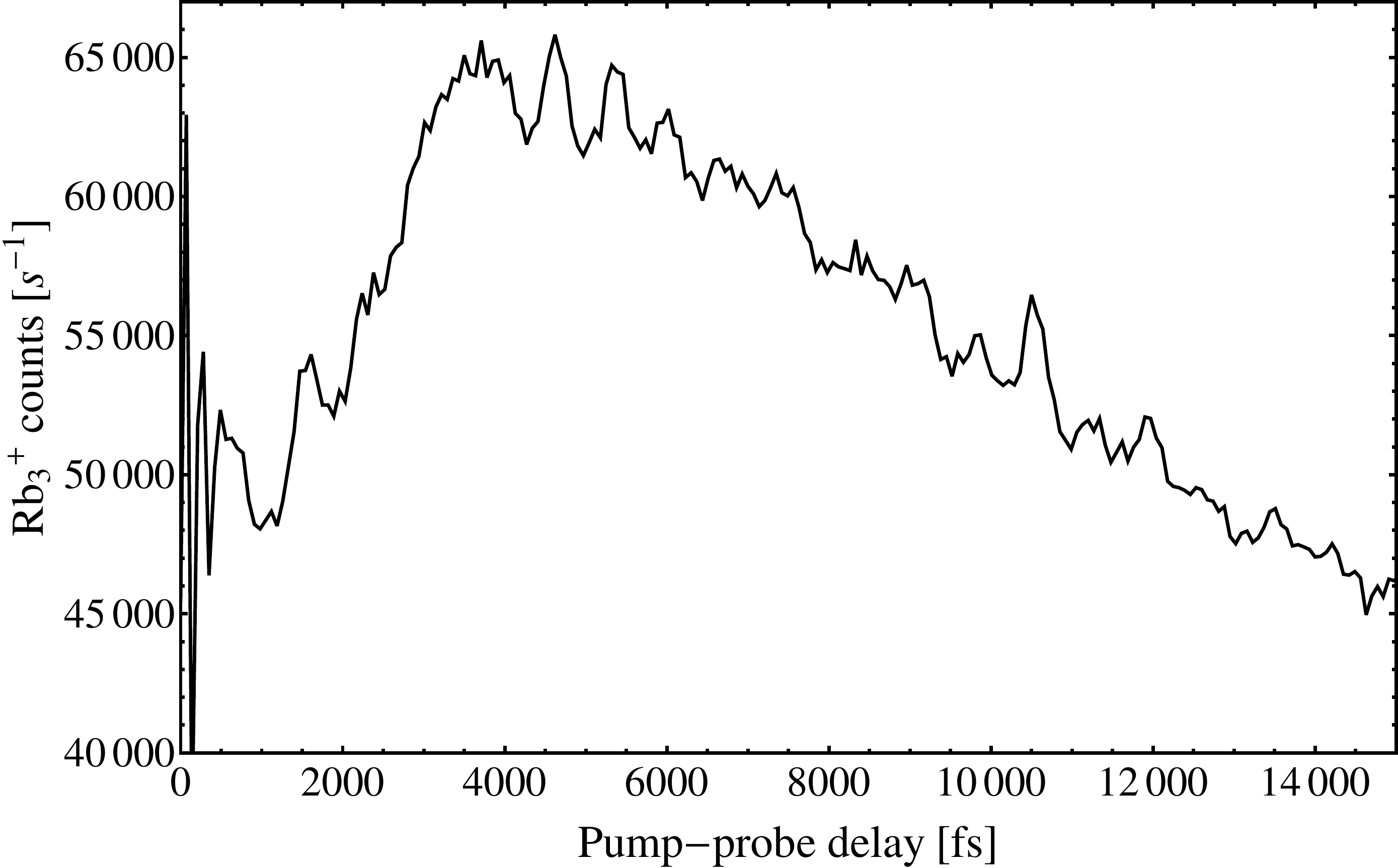}
  \caption{Typical pump-probe photoionization transient recorded at
    the mass of Rb$_3$ at a laser wavelength
     of 850\,\unit{nm} (11765\wn{}). The oscillatory pump-probe
    signal is superimposed by a slowly varying offset resulting from
    photo-fragmentation.}
\label{fig:PPtransient}
\end{center}
\end{figure}

In a first step, photoionization (PI) spectra using single laser
pulses are recorded for all observed trimer species in order to
explore the excitation bands and to find suitable wavelength regions
for pump-probe measurements. The PI spectrum of rubidium trimers
Rb$_3$ is shown in Fig.~\ref{fig:PIRb3} for illustration. Three main
features around $11500\,$cm$^{-1}$, $13000\,$cm$^{-1}$, and
$14000\,$cm$^{-1}$ can be assigned to resonant transitions
$2^4$E$^{\prime}\leftarrow 1^4$A$_2^{\prime}$,
$2^4$E$^{\prime}\leftarrow 1^4$A$_2^{\prime{}\prime{}}$, and to
$1^4$A$_{1,2}^{\prime{}\prime{}}\leftarrow 1^4$A$_2^{\prime}$,
respectively.\cite{shell-model} The transition
$2^4$E$^{\prime}\leftarrow 1^4$A$_2^{\prime}$ has been extensively
studied by laser-induced fluorescence (LIF) and magnetically-induced
circular dichroism (MCD)
spectroscopy.\cite{Nagl_jcp:2008,auboeck08,hauser10jcp} Pump-probe
transients have been recorded at all laser wavelengths at which the
photoionization yield is maximum but clear WP oscillations are
observed only in the frequency interval $11500$--$11900$\wn{}
(indicated as shaded area in Fig.~\ref{fig:PIRb3}). The corresponding
pump-probe transients are discussed in detail in
Sec.~\ref{sec:analysis}.

Let us exemplify the analysis of the pump-probe data with the help of
the Rb$_3$-transient recorded at $\lambda =850\,$nm (11765\wn{})
depicted in Fig.~\ref{fig:PPtransient}. It is composed of two parts:
The oscillatory pump-probe signal contains the information about the
vibrational WP dynamics. In addition, the ion yield slowly increases
within the first few picoseconds before decreasing on a longer
timescale. The initial increase of the ion signal level results from
photo-fragmentation of larger clusters into the trimer channel which
does not contribute to the coherent WP
dynamics.\cite{BaumertZPD:1993,photodissKclusRuff,Schreiber} The subsequent dissociation
of the trimers into dimers and atoms leads to the drop of signal level
at later delay times. Similar photo-fragmentation dynamics of small
alkali clusters in low-spin states in the gas-phase have been studied
in detail.\cite{Ruppe:1996,Schreiber} By fitting an exponential decay
model to the falling edge of the transient data, a dissociation time
constant of $\tau_d=8.9(08)\,$\unit{ps} can be inferred, somewhat
longer than the dissociation times $\tau_d = 4$--$7\,$\unit{ps} found for
low-spin K$_3$ in the B-state.\cite{Ruppe:1996}

\begin{figure}[ht]
  \begin{center}
    \includegraphics[width=0.45\textwidth]{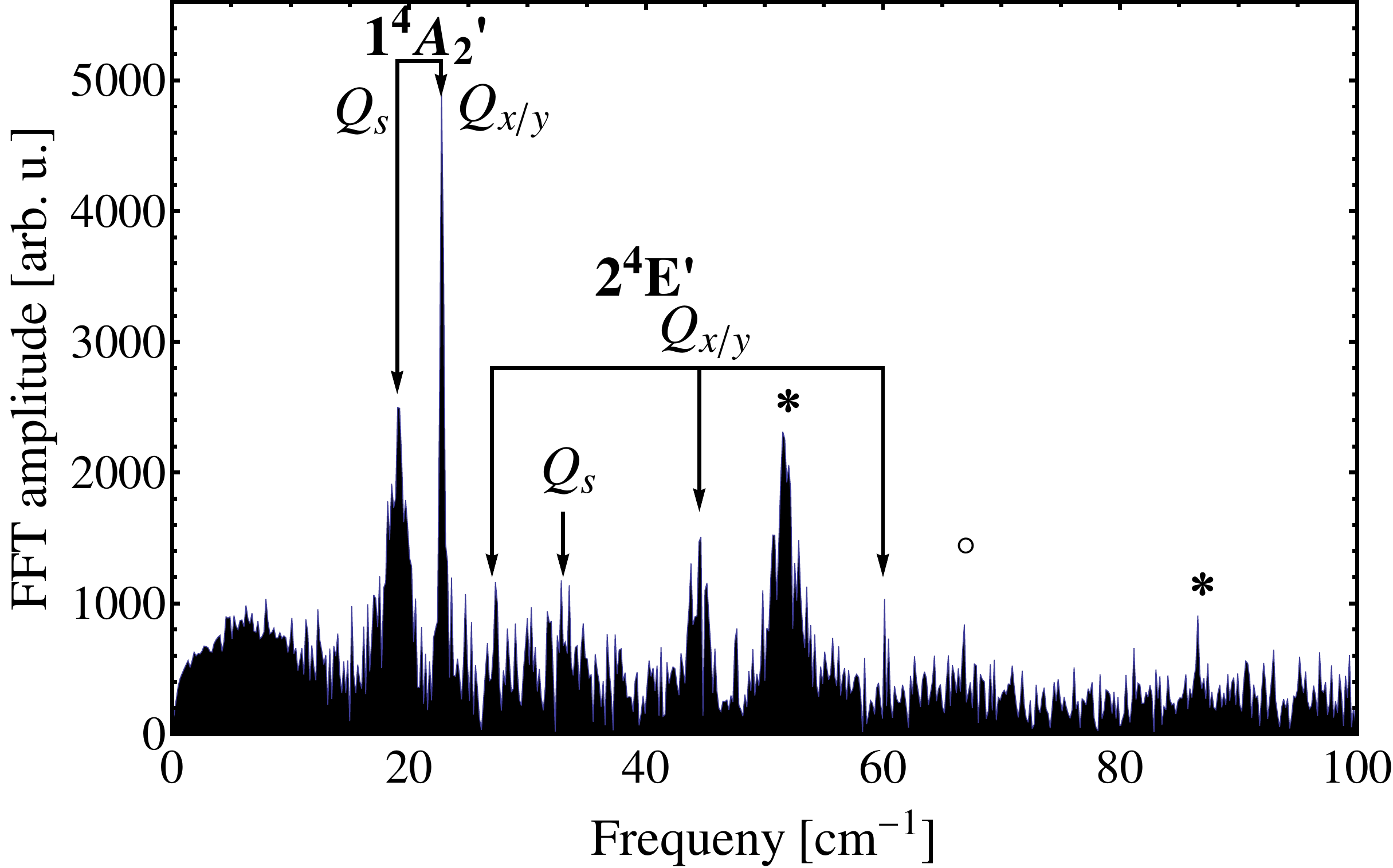}
    \caption{Power spectrum of the Rb$_3$ pump-probe photoionization
      transient recorded at the laser wavelength 845\,\unit{nm}
      (11834\wn{}). Unassigned lines are marked with a star and
      overtone frequencies with a circle.}
\label{fig:FTRb3}
\end{center}
\end{figure}

In order to extract the involved vibrational frequencies, the data are
Fourier transformed after removing the slowly varying envelope using a
high-pass filter (15\wn{} cutoff frequency). This filtering also
removes other longterm signal variations not related to WP-dynamics,
mainly drifts of room and nozzle temperature. We therefore do not
attempt to assign expected low-frequency components. The resulting
power spectrum of Rb$_3$ recorded at 850\,\unit{nm} (11765\wn{}),
depicted in Fig.~\ref{fig:FTRb3}, contains a few clearly discernable
lines. Line positions, widths and amplitudes are analyzed in
Sec.~\ref{sec:analysis} and are assigned to vibrational modes by
comparing with model calculations (Sec.~\ref{sec:theory}). Unassigned
lines are marked with stars, and overtone frequency of assigned lines
with circles.

\begin{figure}[ht]
\begin{center}
  \includegraphics[width=0.45\textwidth]{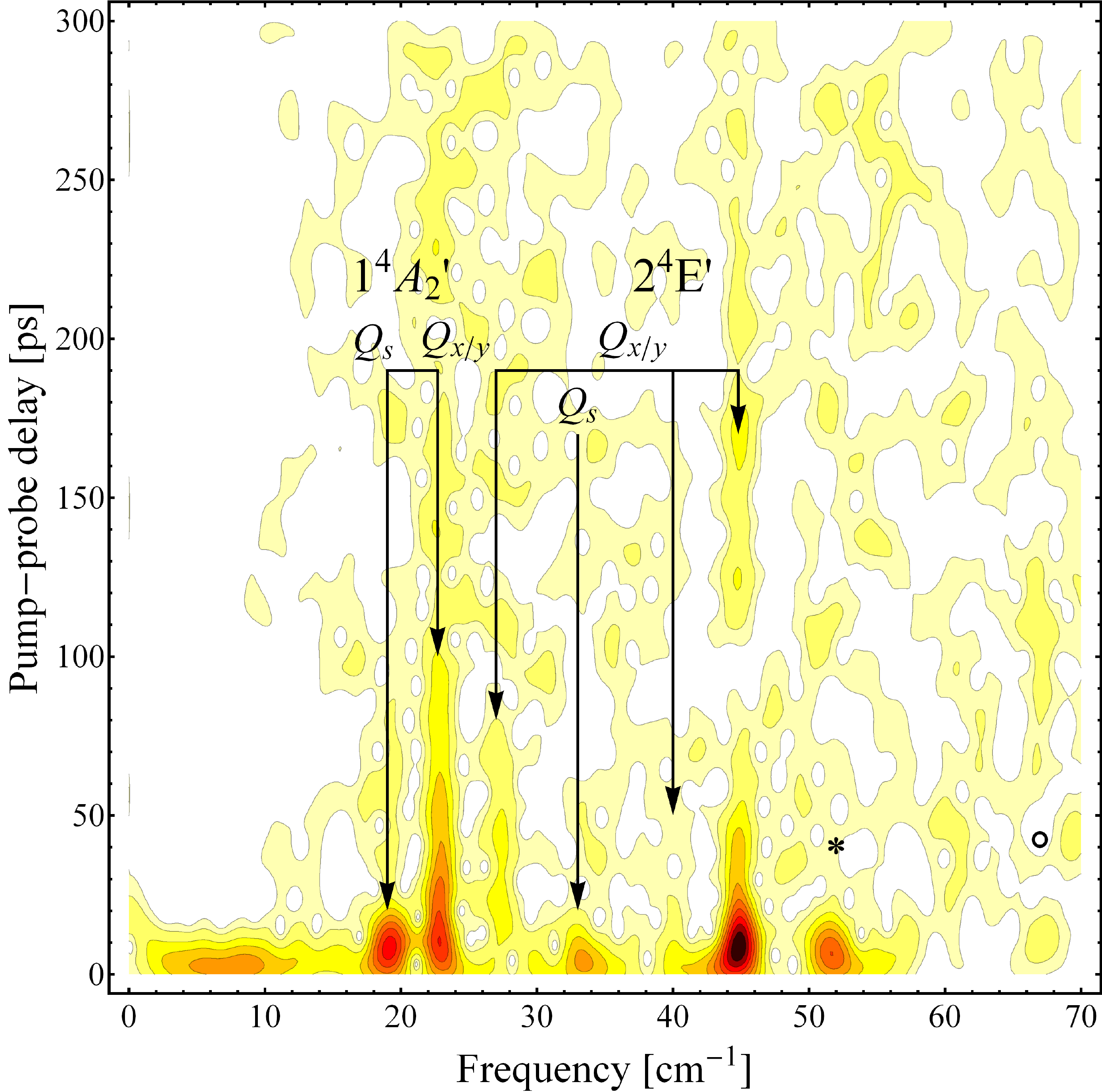}
  \caption{Sliding window Fourier analysis of Rb$_3$ trimers at
    850\,nm (11765\wn). The symmetric stretch modes Q$_s$ are
    subjected to faster dephasing than the asymmetric modes
    Q$_{x/y}$. Overtone fre\-quen\-cies are marked with a circle,
    unassigned lines with a star.}
\label{fig:spekRb850nm}
\end{center}
\end{figure}
As a unique feature of pump-probe spectroscopy, information about the
temporal behavior of the spectral components is accessible. By Fourier
transforming the data using the sliding window technique, so-called
spectrograms can be generated, which contain both spectral as well as
temporal information.\cite{Rutz:1997} Fig.~\ref{fig:spekRb850nm}
displays the spectrogram obtained from Fourier analyzing the
pump-probe data of Rb$_3$ (Fig.~\ref{fig:PPtransient}) using a
Gaussian window function with a full width at half maximum of
15\,\unit{ps} For better visibility the data is plotted with a square
root contour function. The frequency components at about 23\wn{} and
45\wn{}, which are assigned to the asymmetric stretch and bending
modes Q$_{x/y}$ of electronic states $1^4$A$_2^{\prime{}}$ and
$2^4$E$^{\prime}$, respectively, are clearly visible up to delay times
$\Delta t\gtrsim 200\,$ps, whereas the symmetric stretch mode Q$_s$ in the
same states at frequencies 19\wn{} and 33\wn{}, respectively,
disappear within less than 20\,\unit{ps}. The pronounced difference in
dephasing times between symmetric and asymmetric vibrational modes,
which is consistently observed with all other trimer combinations, may
be due to intramolecular vibrational redistribution (IVR) as observed
with low-spin Na$_3$ excited to the B-state.\cite{Rutz:1997,Schreiber}
In order to ascertain this hypothesis quantum dynamics calculations
would have to be performed.\cite{Reischl:1995,Reischl:1996}

Beside this effect, in high-spin trimers attached to helium
nanodroplets, additional dynamics may occur due to quartet-doublet
intersystem crossing as well as to vibrational relaxation induced by
the helium droplet environment.\cite{Higgins:1996,Reho:2001} For
Na$_3$ in the excited $2^4$E$^{\prime{}}$-state the spin-flip reaction
was observed to occur with a time constant ranging between
1.4\,\unit{ns} for the vibrational ground state and 380\,ps for the
asymmetric $v_2=2$-vibration. The probability for the spin-flip
reaction was found to sensitively depend on the spin-orbit coupling
strength. Accordingly, the branching ratio for the occurrence of
unreacted trimers and reaction products is significantly shifted
toward the spin-flip products in the case of K$_3$ as compared to
Na$_3$, suggesting significantly shorter spin-flip times. Thus, even
faster spin-flip dynamics may be expected for Rb$_3$, which features
larger spin-orbit couplings than Na$_3$ and K$_3$. Vibrational
relaxation of the vibration $v=2$ of excited high-spin Na$_3$ occurred
on a timescale of a few nanoseconds. Possibly, the symmetric
`breathing' mode Q$_s$ considered in our experiments couples more
efficiently to the helium environment which would lead to relaxation
times in the picosecond range. Note that cw-spectroscopy of larger
molecules embedded inside helium nanodroplets indicates that vibronic
modes experience variable damping depending on their
symmetry.\cite{Lehnig:2003} The lack of visible revival structures in
Fig.~\ref{fig:spekRb850nm} presumably originates from strongly
anharmonic potential energy surfaces in combination with limited
observation times due to dephasing.

The time-evolution of individual frequency components is extracted
from the spectrogram by applying vertical cuts at the line positions
and then fitting the temporal data with a single or double exponential
decay curve. The corresponding $1/e$-lifetimes $\tau_{1,2}$ are listed in
Tables~\ref{tab:frequRb3} through \ref{tab:frequK3}. Due to the
limited quality of our data a more sophisticated analysis of the
temporal behavior of vibrational modes, which would give insight into
the nature of the dephasing process, is not possible.


\begin{figure}[htb!]
\begin{center}
\includegraphics[width=0.45\textwidth]{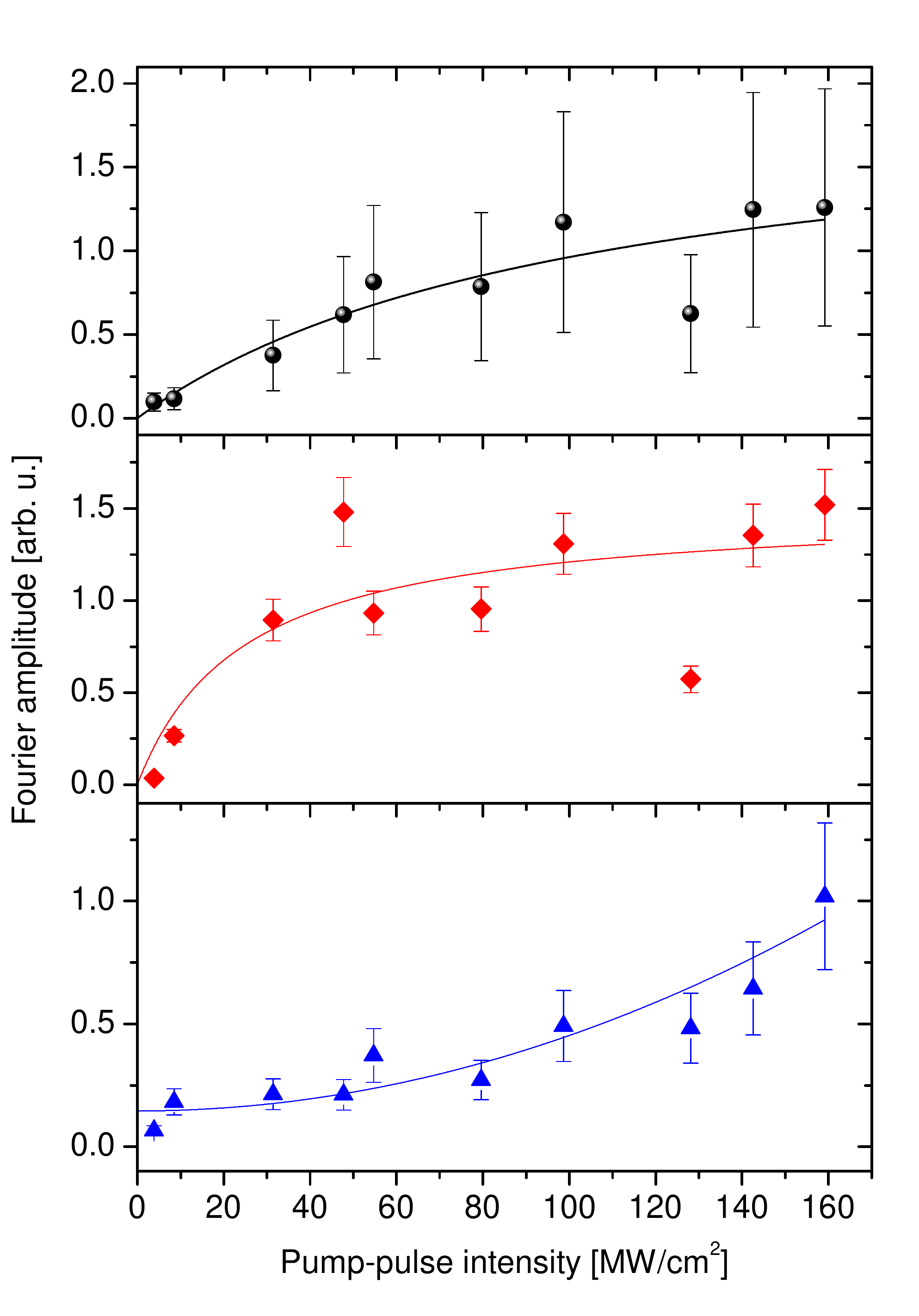}
\caption{Fourier amplitudes for different Rb$_3$ frequency components
  as a function of the pump pulse intensity: upper) Q$_{x/y}$ mode of the
  $1^{4}$A$_{2}^{\prime{}}$ lowest quartet state (22.7\wn{}); middle)
  Q$_{x/y}$ mode of the excited $2^{4}$E$^{\prime}$-state
  (44.7$\wn{}$); lower) unknown component from higher potential
  (52$\wn{}$).}
\label{fig:intplot}
\end{center}
\end{figure}

Additional information for identifying the electronic states involved
in the vibrational dynamics is obtained from analyzing the dependence
of the spectral lines on the pump pulse intensity. In the resonant
multiphoton excitation and ionization schemes exploited in this study,
the amplitudes of spectral components corresponding to vibrations in
different electronic states are expected to feature different scaling
behavior as a function of excitation laser intensity according to the
order of the transition, \textit{i\,e.} the number of photons
involved. In the experiment, the pump-pulse intensity is varied by
placing neutral density filters into the pump beam without changing
the position and size of the laser-droplet interaction volume. In
order to isolate the pure effect of the pump pulse in the pump-probe
scheme, the background count rates, which result from direct 3PI when
applying either only the pump or only the probe pulse, are subtracted
from the data.

The resulting intensity dependence of the line amplitude at about
23\,$\wn$ (Q$_{x/y}$ modes of Rb$_3$ in the $1^{4}$A$_{2}^{\prime}$
lowest quartet state), shown in Fig.~\ref{fig:intplot}a, is nearly
linear, as opposed to clearly non-linear saturation behavior observed
for the component at about 45$\wn$ (Q$_{x/y}$ mode of the excited
$2^{4}$E$^{\prime}$ state), shown in Fig.~\ref{fig:intplot}b. This is
consistent with the expected lower probability for creating a WP in
the quartet ground state $1^{4}$A$_{2}^{\prime}$, which requires a
two-photon RISRS-process, as compared to single-photon excitation to
the $2^{4}$E$^{\prime}$-state. The unassigned broad feature at
52$\wn$, however, shows an unsaturated non-linearly rising intensity
dependence which points at a multiphoton transition to a higher lying
state being active. The solid lines represent simple model fit curves
to guide the eye.

\begin{figure}[ht]
\begin{center}
\includegraphics[width=0.45\textwidth]{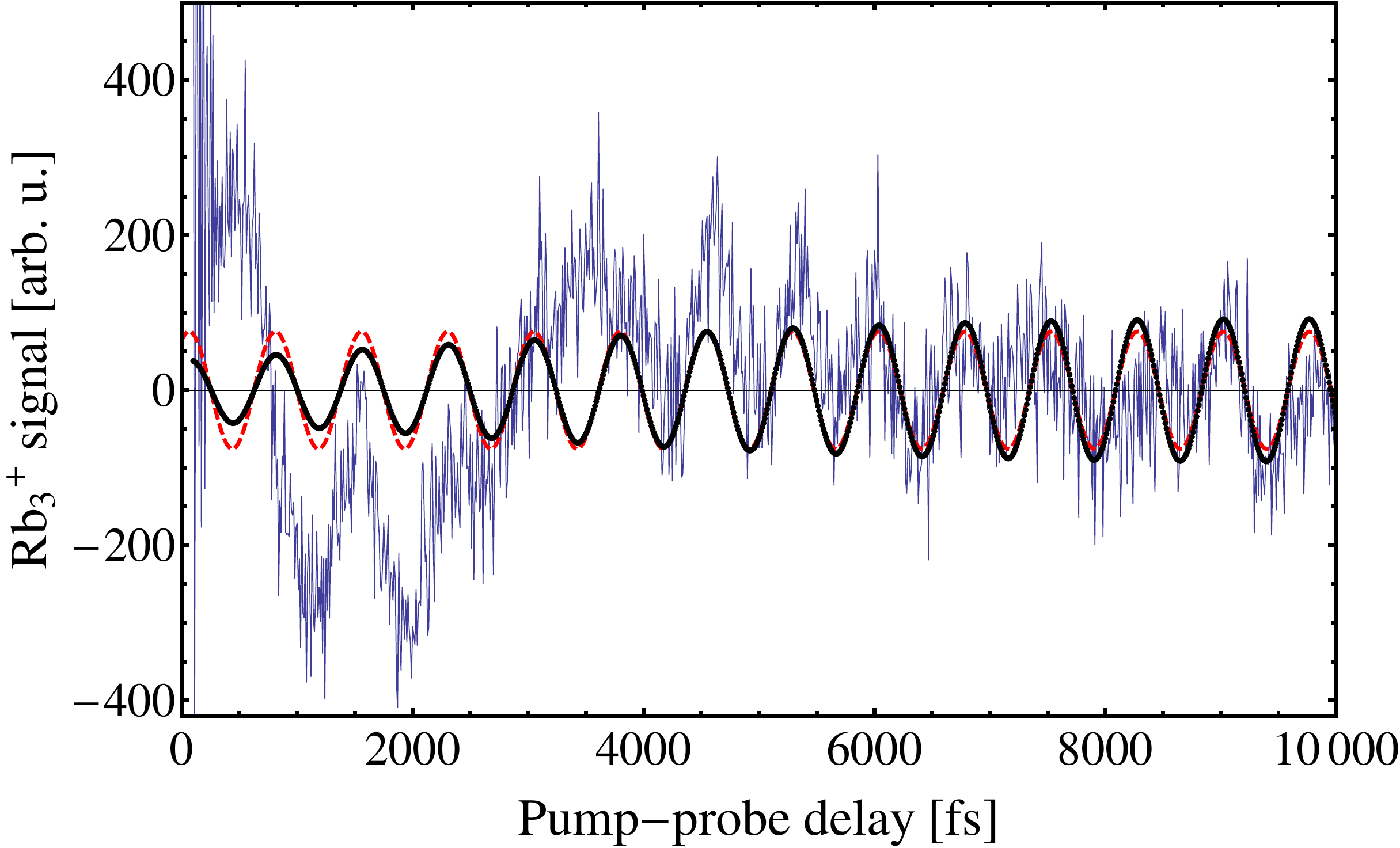}
\caption{Transient signal of Rb$_3$ at 850\,\unit{nm}. The black line is
  obtained by notch filtering the measured total transient at
  44$\wn$. The initial phase is determined from a sinusodial fit
  (dashed, red line).}
\label{fig:phases}
\end{center}
\end{figure}
Furthermore, information about the pathway along which the probability
density propagates on its way from the initial vibronic ground state
up to the final ionic state can be obtained from analyzing the initial
phases $\Phi$ of the individual oscillation components. For the
example of a diatomic molecule, when a WP is excited and probed at the
same classical turning point of a potential energy curve one expects
the phase to be zero, whereas a phase of $\Phi\approx \pi$ means that
the WP is created at one turning point and probed at the other, as
observed for the excited state dynamics of K$_2$ and Rb$_2$ formed on
helium nanodroplets.\cite{Claas:2006,Mudrich:2009} The ground state
vibrations induced by RISRS in triplet Rb$_2$ were found to have
initial phases $\Phi\approx 0.3\,2\pi$ and $\Phi\approx 0.7\, 2\pi$
depending on the ionization path which differed from those of the
excited state oscillations and thus helped to discern ground and
excited state dynamics.\cite{Mudrich:2009}

For the determination of initial phases $\Phi $ in the present
measurements, the data are manipulated as follows. A narrow-band notch
filter is applied for isolating each frequency component. This is done
by multiplying the data in the frequency domain by a narrow gaussian
window function ($1/e^2$-width of 0.5\wn) centered around the
frequency of interest. The inverse Fourier transform yields the
filtered transient data and the initial phase is then obtained from
fitting a sinusoidal function, as illustrated in Fig.~\ref{fig:phases}
for the 44.7$\wn$ component of Rb$_3$ excited at 850\,\unit{nm}
(11765\wn). The resulting initial phases for all prominent lines of
all trimer species are shown in Tables~\ref{tab:frequRb3} through
\ref{tab:frequK3} and are discussed in Section~\ref{sec:analysis}.

\section{Theory}\label{sec:theory}

For the interpretation of vibrationally resolved spectra a good
knowledge of the involved PES is indispensable. In the case of the
homonuclear trimers K$_3$ and Rb$_3$ we fall back to the \emph{ab
  initio} results of our previous articles on the
quartet~\cite{hauser10jcp} and doublet~\cite{hauser10cp} states. For
our approach to the mixed trimers K$_2$Rb and KRb$_2$ we refer the
potential surfaces and the corresponding energy levels of
Paper~I.\cite{paper1} Since the alkali metal trimers on the surface of
helium droplets are preferably formed in the van der Waals bound
high-spin states, we concentrate on the quartet manifold of the
electronic states and pick the experimentally relevant states between
11500 and 12100~cm$^{-1}$.  We find that due to the similar electronic
state structure of all trimer species the same pair of states, namely
3$^4$A$_1$ and 4$^4$B$_2$, is involved in the electronic
excitation. However, for the homonuclear trimers we will use the
labels of the D$_{3h}$ point group because of the increased symmetry
at equilateral triangle geometry, where these pairs of states form a
doubly degenerate 2$^4$E$^{\prime}$ state. The lowest quartet states
may be labeled as 1$^4$B$_2$ in C$_{2v}$, corresponding to
1$^4$A$_2^{\prime}$ states in D$_{3h}$. We will discuss the vibronic
spectra predictions for the homo- and hetereonuclear trimers in
Sections~\ref{rb3}-\,\ref{k3}.

\begin{figure}[htb]
\begin{center}
\includegraphics[width=0.45\textwidth]{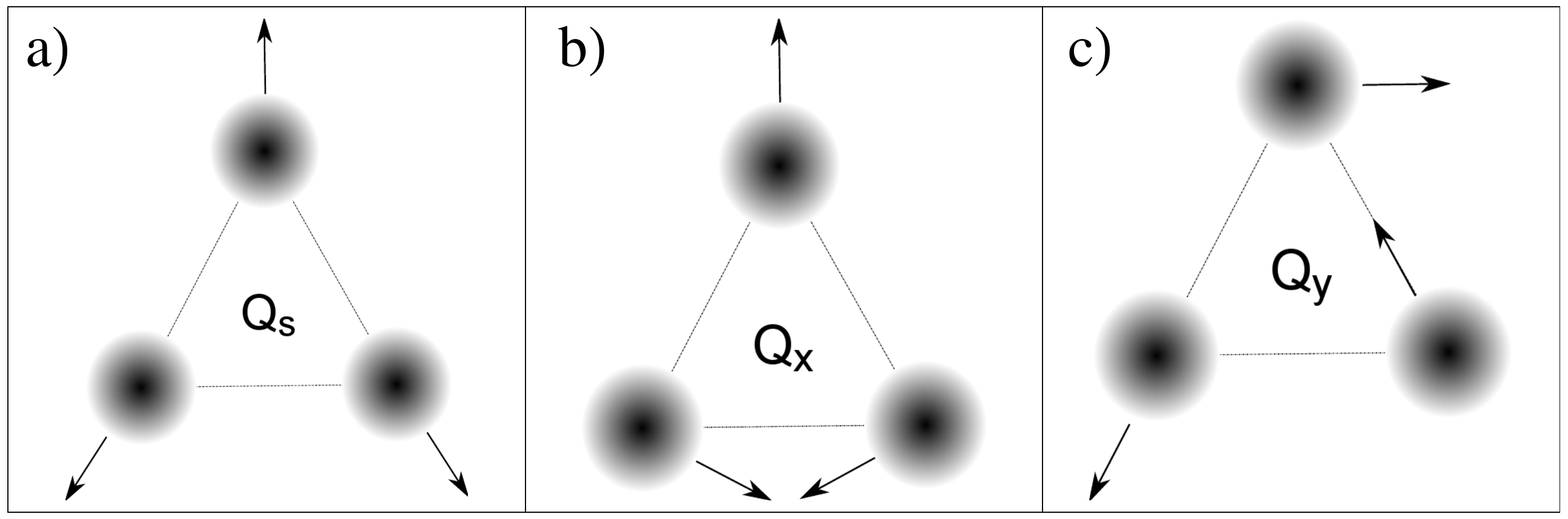}
\caption{Sketched vibrational modes of a homonuclear trimer:
  (a) symmetric stretch mode, (b) asymmetric bending mode,
  (c) asymmetric stretch mode.}
\label{fig:trimermodes}
\end{center}
\end{figure}

In the case of the highly symmetric equilateral homonuclear trimers
the three normal modes Q$_s$ (breathing mode), Q$_x$ (symmetric mode)
and Q$_y$ (asymmetric mode) can be obtained -- except for a constant
factor -- directly from symmetry considerations. They are shown in
Figure~\ref{fig:trimermodes}.

For the heteronuclear trimers, however, they have to be calculated
numerically as described in Paper~I.\cite{paper1} Note that the Q$_x$
and Q$_y$ modes are degenerate in the homonuclear trimers, giving rise
to a Jahn-Teller distortion in electronically degenerate states. The
vibrational degeneracy is lifted by a few wavenumbers in the case of
the heteronuclear trimers.

\begin{figure*}[htb]
\begin{center}
\includegraphics[width=1.0\textwidth]{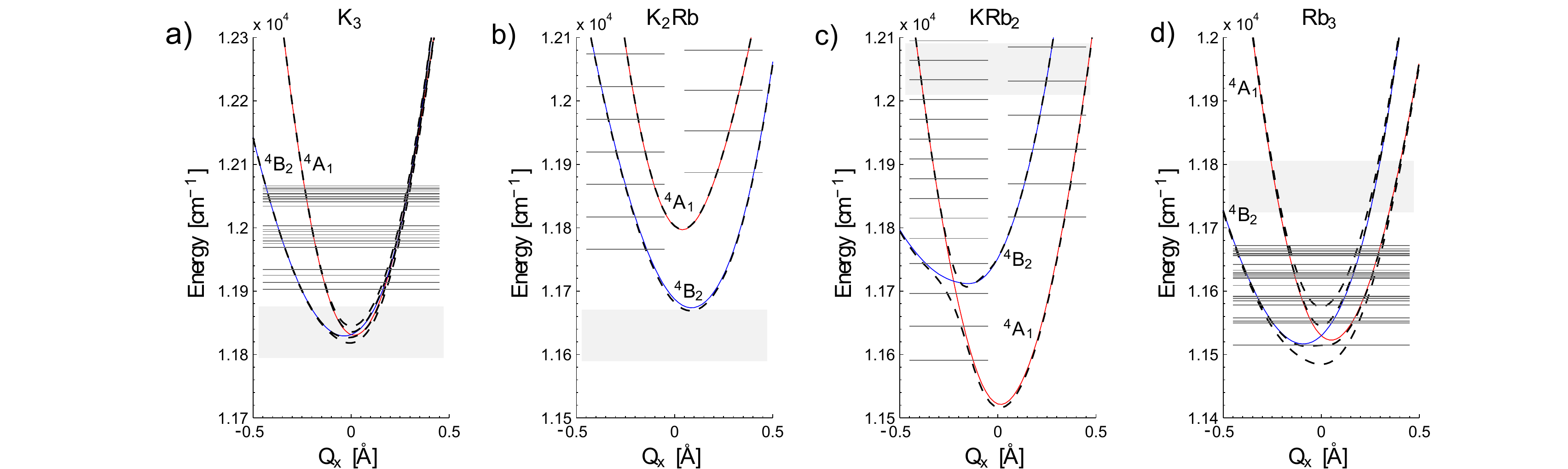}
\caption{Subfigures a)-d) show cuts through the PES of the involved
  3$^4$A$_1$ (red) and 4$^4$B$_2$ (blue) electronically excited states
  as a function of Q$_x$ for all investigated alkali
  trimers. SO-corrected surfaces are plotted as black-dashed
  lines. The first few vibronic levels of all SO-PES are indicated by
  dark grey lines. Experimental laser excitation windows (FWHM of
  80~cm$^{-1}$) of prominent power spectra (see
  Figs. \ref{fig:FTRb3},\ref{fig:FTKRb2}, \ref{fig:FTK2Rb} and
  \ref{fig:FTK3}) are plotted as shaded areas.  The energy scale is
  referenced to the vibrational ground state of the $1^4$B$_2$ lowest
  quartet PES.}
\label{fig:qx-all}
\end{center}
\end{figure*}

\subsection{The 2$^4$E$^{\prime}\leftarrow{}1^4$A$_2^{\prime}$
  transitions in K$_3$ and Rb$_3$}
\label{sec:involv-electr-1}
From previous calculations the electronically excited
2$^4$E$^{\prime}$ states of K$_3$ and Rb$_3$ with vertical excitation
energies of 11855 and 11550~cm$^{-1}$, respectively, could be
identified as the states of major interest for the experimental
femtosecond-laser excitation of the homonuclear trimers. In both
molecules the 2$^4$E$^{\prime}$ state is affected by a combination of
Jahn-Teller distortion and spin-orbit coupling. This can be seen in
Figures~\ref{fig:qx-all}a and~\ref{fig:qx-all}d, where the involved
PES for K$_3$ and Rb$_3$ are plotted as function of Q$_x$. The two
branches of the non-relativistic PES can be labeled as 3$^4$A$_1$ and
4$^4$B$_2$ in the C$_{2v}$ point group. Inclusion of SO-coupling leads
to a new system of four PES shown as black-dashed lines. In
Ref.~\citenum{hauser10jcp} we provided a detailed analysis of these
states in terms of the relativistic Jahn-Teller effect theory and
generated spectral simulations for the interpretation of vibrationally
resolved experimental data based on a combination of LIF spectroscopy
and MCD techniques. For the interpretation of the femtosecond
pump-probe data we fall back to the previously calculated vibrational
eigenstates in the JT-distored potential energy surfaces and list them
in Tabs.~\ref{tab:eig-rb} and~\ref{tab:eig-k} for Rb$_3$ and K$_3$,
respectively.
The pump-probe Fourier-spectra are expected to feature difference
frequencies between these levels that we will represent as $(n,m)$ for
a beat frequency of levels $n$ and $m$. The most promising candidates
in terms of matching frequency differences as well as excitation
probabilities are printed in bold.  Vibrational excitations of the
slightly anharmonic Q$_s$ mode ($\nu_{Q_s}\approx53$\wn for K$_3$ and
33\wn for Rb$_3$) are not listed here. The intensities shown in the
tables are based on Franck-Condon integrals for transitions from the
$v=0$ state of the 1$^4$A$^{\prime}_2$ lowest quartet state. This
information is only weakly related to the much more complex
experimental situation of coherent multi-level coupling of different
electronically excited states, but may be seen as a zeroth order
criteria for the selection of vibronic levels involved. The vibronic
levels are also shown in Figure~\ref{fig:qx-all} as dark gray
lines. Note the high density of levels for the homonuclear trimers
caused by SO-splitting.

\subsection{The 3$^4$A$_{1},\,$4$^4$B$_{2}\leftarrow{}$1$^4$B$_2$
  transitions in K$_2$Rb and KRb$_2$}
\label{sec:involv-electr-2}
For the mixed trimers the excited state structure of the quartet
manifolds is strongly related to the structure of the homonuclear
species in terms of vertical energies and state ordering. This can be
seen in Figures~\ref{fig:qx-all}b and~\ref{fig:qx-all}c for K$_2$Rb
and KRb$_2$, respectively. The same pairs of states, namely 3$^4$A$_1$
and 4$^4$B$_2$, are involved in the experiment. However, due to the
reduction of the molecular symmetry the vibrational degeneracy of
Q$_x$ and Q$_y$ is lifted and the JT-coupling is
suppressed. Nevertheless, the electronic degeneracy remains. In the
case of KRb$_2$ the intersection of the two electronic states lies in
the experimentally relevant energy range as can be seen in
Figure~\ref{fig:qx-all}c. The vertical excitation energies are
11429~cm$^{-1}$ and 11659~cm$^{-1}$ for the 3$^4$A$_1$ and the
4$^4$B$_2$ states of KRb$_2$, respectively. In the case of K$_2$Rb the
same states are found at 11691~cm$^{-1}$ and 11577~cm$^{-1}$,
respectively. Following the argumentation given in
Paper~I\cite{paper1} the SO coupling can be neglected in the case of
the heteronuclear trimers except for the fact that the conical
intersection of the two states as a function of Q$_x$ becomes an
avoided crossing. However, this leads, especially for KRb$_2$, to a
different, strongly anharmonic PES structure with the corresponding
vibronic levels listed in Tables~2 and~3 of Paper~I.\cite{paper1}

\section{Fourier spectra analysis}
\label{sec:analysis}

In this section we compare the measured Fourier spectra with
frequencies derived from the \emph{ab initio} calculations. In most
cases harmonic approximations to the given potential surfaces are
sufficient for a qualitative peak assignment. Therefore, when not
specified, given calculated values are the harmonic frequencies to the
concerned PES. Exceptions are the Jahn-Teller distorted pairs of
excited states in the homonuclear trimers Rb$_3$ and K$_3$, the
excited states of KRb$_2$ due to the avoided crossing in Q$_x$ and the
slightly anharmonic Q$_s$ modes of all trimers. In the first case we
fall back to selected numerical results for the vibronic eigenstates
listed in Tables~\ref{tab:eig-rb} and~\ref{tab:eig-k}. In the second,
we refer to Tables~2 and~3 of Paper~I, where a finite-difference
approach is applied to solve the nuclear part of the Schr\"{o}dinger
equation explicitely for the given SO-corrected \emph{ab initio}
potentials. For the sake of a general overview and a direct comparison
the harmonic frequency approximations of all involved states are
summarized in Table~\ref{tab:freq-sum}. However, note that the actual
structure of vibronic levels (see Tables~\ref{tab:eig-rb}
and~\ref{tab:eig-k}) in the $2^4$E$^{\prime{}}$ states of the
homonuclear trimers deviates significantly from the harmonic
situation, so the harmonic frequencies in Q$_x$ and Q$_y$ are of
little value for further interpretations. Instead, particular
differences between vibrational levels in the excited states are
expected to be present in the Fourier spectra.

\begin{table}[ht]
\small
\caption{Harmonic frequency approximations (in cm$^{-1}$) of all
  involved quartet states K$_2$Rb, KRb$_2$ (C$_{2v}$ point group nomenclature) and
  Rb$_3$ and K$_3$(D$_{3h}$) }
\label{tab:freq-sum}
\begin{tabular*}{0.45\textwidth}{@{\extracolsep{\fill}}lcccccc}
\hline
   &\multicolumn{3}{c}{K$_3$} & \multicolumn{3}{c}{K$_2$Rb}\\
 mode &1$^4$A$_2^{\prime}$ &2$^4$E$^{\prime}$ &  &1$^4$B$_2$ & 3$^4$A$_1$ &4$^4$B$_2$ \\
\hline
Q$_x$ &33 & 67&  &34 & 65& 51\\
Q$_y$ &33 & 67&  &35 &57 & 68\\
Q$_s$ &37 & 52&  &34 &56 & 49\\
\hline
\hline
   &\multicolumn{3}{c}{KRb$_2$} & \multicolumn{3}{c}{Rb$_3$}\\
 mode &1$^4$B$_2$ &3$^4$A$_1$ &4$^4$B$_2$ &1$^4$A$_2^{\prime}$ &2$^4$E$^{\prime}$ & \\
\hline
Q$_x$ &26 &54 &52 &21 & 42& \\
Q$_y$ &31 &51 &62 &21 & 42& \\
Q$_s$ &28 &46 &52 &19 & 33& \\
\hline
\end{tabular*}
\end{table}

\subsection{The Rb$_3$ trimer}
\label{rb3}
The rubidium trimer Rb$_3$ is the heaviest of the observed species
with an average mass of $m=255.5$~amu. WP oscillation signals are
recorded at different laser wavelengths between about $840\,$nm
($11900$\wn{}) and $870\,$nm ($11500$\wn{})
(Fig.~\ref{fig:PPtransient}).  The obtained Fourier spectra show
multiple vibrational lines with different amplitude and width most of
which are visible in a window of about 10\,\unit{nm} around $850\,$nm
($11765$\wn{}) (Fig.~\ref{fig:FTRb3}). Naively, one would expect a
total of 4 lines in the case of the homonuclear trimers: The symmetric
Q$_s$ and the degenerate Q$_{x/y}$ modes in the lowest quartet
$1^{4}$A$_{2}^{\prime}$ and in the excited quartet $2^{4}$E$^{\prime}$
states.

All relevant spectral information about Rb$_3$ is compiled in
Tab.~\ref{tab:frequRb3}.  For the $1^{4}$A$_{2}^{\prime}$ state the
fitted harmonic frequencies of the Q$_s$ ($19$\wn) and the
Q$_{x/y}$-modes ($21$\wn) are in good agreement with the experimental
values, $19.1$\wn{} and $22.7$\wn{}. The excited $2^{4}$E$^{\prime}$
state shows strong Jahn-Teller and SO couplings which requires a
correction of the PES as well as the numerical calculation of vibronic
eigenstates for the coupled Q$_{x/y}$ modes.\cite{hauser10jcp} As can
be seen in Figure~\ref{fig:qx-all}, at 850~nm (11765\wn{}) the laser
excitation lies about 200\wn{} above the conical intersection, so
significant contributions of several excited vibronic levels are
expected.  Levels of particular interest are those with reasonable FC
overlap with the vibrational ground state of the $1^4$A$_2^{\prime{}}$
lowest quartet state.

\begin{table}[htb!]
\small
\begin{center}
  \caption{Positions of vibronic levels (in cm$^{-1}$) in the
    JT-distorted 2$^4$E$^{\prime}$ electronically excited state of
    Rb$_3$. The harmonic frequency of the degenerate Q$_{x/y}$ mode is
    42~cm$^{-1}$.\label{tab:eig-rb}}
\begin{tabular*}{0.45\textwidth}{@{\extracolsep{\fill}}cccc} \hline
  Nr.&abs. position    &rel. position$^a$ & intensity \\ \hline
\textbf{1}     &    \textbf{11515}     &     \textbf{-35}   &       \textbf{1.42}  \\
2     &    11550    &        0   &       0.09  \\
3     &    11553    &        3   &       0.00  \\
\textbf{4}     &    \textbf{11558}    &        \textbf{8}   &       \textbf{1.14}  \\
\textbf{5}     &    \textbf{11578}    &       \textbf{28}   &     \textbf{0.47}  \\
6     &    11585    &       35   &       0.00  \\
7     &    11588    &       38   &       0.16  \\
8     &    11591    &       41   &       0.00  \\
9     &    11592    &       42   &       0.00  \\
0     &    11609    &       59   &       0.00  \\
\textbf{10}    &    \textbf{11620}    &       \textbf{70}   &     \textbf{0.17}  \\
11    &    11621    &       71   &       0.00  \\
12    &    11624    &       74   &       0.02  \\
13    &    11627    &       77   &       0.00  \\
14    &    11629    &       79   &       0.00  \\
\textbf{15}    &    \textbf{11633}    &       \textbf{83}   &     \textbf{0.70}  \\
16    &    11642    &       92   &       0.00  \\
\textbf{17}    &    \textbf{11656}    &      \textbf{106}   &       \textbf{0.16}  \\
18    &    11657    &      107   &       0.00  \\
19    &    11659    &      109   &       0.00  \\
20    &    11660    &      110   &       0.00  \\
21    &    11663    &      113   &       0.00  \\
22    &    11665    &      115   &       0.00  \\
23    &    11667    &      117   &       0.00  \\
\textbf{24}    &    \textbf{11672}    &      \textbf{122}   &    \textbf{1.36}  \\ \hline
     	\end{tabular*}		
\end{center}
$^a$Energetic difference to the point of the conical intersection, assuming no
zero-point energy in Q$_s$.
\end{table}

According to the theoretical data listed in Tab.~\ref{tab:eig-rb} the
vibrational frequency beats (1,4) (43\wn{}) and (5,10) (42\wn{}) are
in fair agreement with the experimental peak at 44.7\wn{}. Note that
-- accidentally-- this value is also close to the hypothetical value
(42\wn{}) predicted by Jahn-Teller effect theory. However, the full
treatment of the problem includes SO-coupling and leads to 4 potential
curves with spin projections $\Sigma=3/2,1/2,-1/2,-3/2$. The states 1
and 4 constituing the (1,4) beat have different spin character. In the
case of the $2^{4}$E$^{\prime}$-states of Rb$_3$ as well as for K$_3$
the calculated states in Tables~\ref{tab:eig-rb} and \ref{tab:eig-k}
are just labeled with numbers to refer to their absolute energy and do
not represent vibrational quantum numbers.  Note that the observed
frequency also fits well to a $\Delta v=2$ overtone beat of the
Q$_{x/y}$-mode of the $1^4$A$_2^{\prime{}}$ PES (2 22.7\wn). This
component is observed over a broader band of excitation wavelengths
(500\wn) than in the cw excitation spectrum (300\wn{}) of the same
system\cite{nagl:prl08}. This supports the interpretation in terms of
a Raman transition to the lowest quartet state $1^4$A$_2^{\prime{}}$.
However, the combined large laser bandwidth (80\wn{}) and line
broadening due to the high laser intensity can also explain the
broadened spectrum with respect to the cw case. Furthermore, it is
surprising, that the overtone should be more intensive than the
fundamental beat component, which is the case for many of the recorded
spectra at other wavelengths not shown in this publication.

The slightly anharmonic symmetric stretch mode Q$_s$ of the
$2^{4}$E$^{\prime}$-state is analyzed by integrating the
one-dimensional Schr\"{o}dinger equation. The frequency difference
between the lowest vibrational quanta $v=0$ and $v=1$, (33\,cm$^{-1}$)
is in very good agreement with the experimental value
33.4\wn{}. Generally, the FC factors for the RISRS excitation seem to
favor the population of the lowest $v$-states. A comparison of the
absolute line strengths was not attempted since the higher lying PES
involved in the REMPI probing step are yet unknown.

Besides the assigned lines we find a few more peaks in the Fourier
spectrum of Rb$_3$. The quite intense line at $52$\wn{} is attributed
to the WP dynamics in a higher lying electronic state that is excited
by multiphoton absorption. This interpretation is based on the
nonlinear scaling of the line amplitude as a function of pump-laser
intensity discussed in Sec.~\ref{sec:ppspectra}. Moreover, the initial
phase $\Phi=0.67(5)\, 2 \pi$ of the WP oscillation at 52\wn
significantly differs from $\Phi$ of the vibrational transients of the
$1^4$A$_2^{\prime{}}$ and $2^4$E$^{\prime{}}$-states. The latter are
consistent with $\Phi =0$, indicating coinciding FC-regions for the
excitation and the ionization transitions. The weak line at
$66.8$\wn{} may be identified as the first overtone frequency of the
Q$_s$-mode in the $2^4$E$^{\prime}$-state ($33$\wn{}). Following the
suggested approach of picking vibronic levels that have reasonable FC
overlap with the lowest vibrational eigenfunction of the
$1^{4}$A$_{2}^{\prime}$ state one might relate the difference
frequencies (15,17) ($30$\wn{}) and (15,24) ($39$\wn{}) extracted from
Table~\ref{tab:eig-rb} to the weak beats at 27 and 40\wn{},
respectively (visible in Fig.~\ref{fig:spekRb850nm}).  Likewise, the
weak frequency component at 60\wn{} could be related to the calculated
frequency (1,5) (63\wn), whereas there is no match for the observed
frequency at 86.6\wn{}.

The fast decay of the Q$_{s}$ modes, visualized in
Fig.~\ref{fig:spekRb850nm}, is consistent with the lifetimes obtained
from the line widths of the integral Fourier spectra. As discussed in
Sec.~\ref{sec:ppspectra}, several ultrafast molecular processes are
conceivable, including intramolecular vibrational redistribution (IVR)
as observed in the
gas-phase,\cite{RutzZPD:1997,Schreiber,Reischl:1995,Reischl:1996}
intersystem-crossing or vibrational relaxation induced by the helium
droplets, as observed with helium droplet-formed alkali
trimers.\cite{Higgins:1996,Reho:2001} The fact that the Q$_s$ mode has
not been observed in cw spectra of Na$_3$ in the
gas-phase\cite{HerrmanncwNa3,pseudorotNa3Ernst} nor with mixed
rubidium-potassium trimers formed on helium
nanodroplets\cite{nagl:prl08} points at a fast intramolecular process,
presumably IVR, to cause the fast decay of the Q$_s$-signal.

The Q$_{x/y}$ mode has a much longer lifetime $\sim 100$\unit{ps},
possibly limited by quartet-doublet intersystem
crossing.\cite{Higgins:1996,Reho:2001}
This explains why the cw spectra are entirely dominated by the
Q$_{x/y}$ mode of the excited states. Interestingly, the degenerate
mode shows a double exponential decay with two very different time
constants. While most of the intensity is lost within a few
picoseconds after the excitation, the second decay lasts for hundreds
of picoseconds. This leads to a bimodal spectral line structure with a
broader base and a nearly Fourier limited central part. This effect
could be related to strong initial trimer-droplet interactions
followed by the reorganisation of the helium surrounding or
desorption of the molecules from the helium droplet
surface. Alkali atoms are known to leave the droplets upon electronic
excitation and the observed fast decay time is in rough agreement with
expected desorption times of about 10\unit{ps}.\cite{Takayanagi:2003}
The exact value may depend on the dopant species and on the state the
dopant is excited to.\cite{Reho:2001,Auboeck:2008,Gruner:2011}
Unfortunately, the quality of the recorded data does not allow a more
detailed analysis. Future two-color pump-probe experiments will
improve the signal quality by choosing appropriate probe frequencies
for single-photon PI.

\begin{table}[ht]
\small
\begin{center}
  \caption{Measured vibrational beat frequencies and calculated
    frequencies for different vibrational modes in two electronic
    states of Rb$_3$. The most prominent frequencies are marked in bold.}
  \begin{tabular}{p{0.9cm}p{0.9cm}p{0.9cm}p{0.9cm}p{1.3cm}p{0.9cm}}\hline
    $\nu_{exp}$     & $\tau_1$    & $\tau_2$       & $\Phi$         & State		                         &  $\nu_{theo}$  \\
    $[\wn]$         &$[\unit{ps}]$&$[\unit{ps}]$   &$[\unit{2 \, \pi}]$  & (mode)                           &  $[\wn]$\\
    \hline
    \textbf{19.1(2)}	        & 16(3)       & -    					 & 0.08(1)        & $1^{4}$A$_{2}^{\prime}($Q$_{s}$)   & 19    \\
    \textbf{22.7(3)}         & 33(15)      &160(110)        & 0.15(6)        & $1^{4}$A$_{2}^{\prime}($Q$_{x/y})$ & 21    \\
    27    			    & -           & -   				   & -  				  	& $2^{4}$E$^{\prime}\,(15,17)$       & 30    \\
    33.4  			    & -   	      & -   					 & -              & $2^{4}$E$^{\prime}($Q$_{s}$)	     & 33    \\
    40 	 			      & -     		  & -					     & -	   			    & $2^{4}$E$^{\prime}\,(15,24)$       & 39    \\
    \textbf{44.7(2)}         & 16(10)			& -              & 0.03(5)        & $2^{4}$E$^{\prime}\,(1,4)$  	     & 42 	 \\
    \textbf{52(2)}           & 9(1)        & -              & 0.67(5)	      & - 		      	  	      				   & - 	   \\
    60              & -           & -              & -      	      & $2^{4}$E$^{\prime}\,(1,5)$  		   & 63 	 \\
    66.8            & -           & -              & -              & $2^{4}$E$^{\prime}($Q$_{s}$)       & $2\times 33$ \\
    86.6						& -						& -						   & -							& $2^{4}$E$^{\prime}\,(1,4)$			   & $2\times 42$ \\

    \hline
\end{tabular}
\label{tab:frequRb3}
\end{center}
\end{table}

\subsection{The KRb$_2$ trimer}
\label{krb2}
\begin{figure}[ht]
  \begin{center}{ 
  \includegraphics[width=0.95\columnwidth]{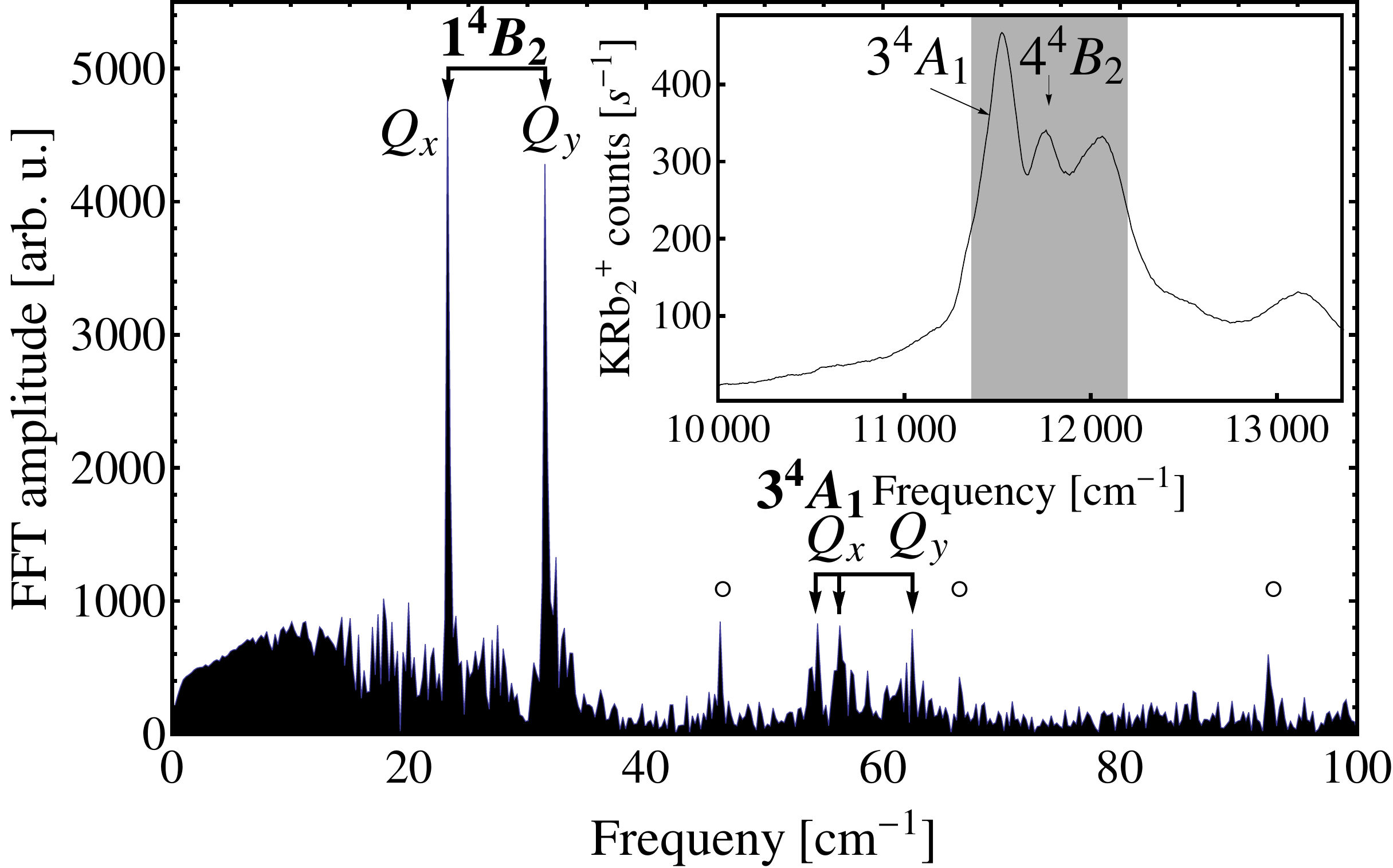}
      \caption{Fourier transform of the KRb$_2$ pump-probe
        photoionization transient recorded at 830\,\unit{nm} (12048\wn{}). The
        corresponding photoionization spectrum is shown in the inset.
        \label{fig:FTKRb2}}
}
\end{center}
\end{figure}

Fourier spectra of KRb$_2$ were recorded in the same way as for Rb$_3$
but with the quadrupole mass spectrometer set to the mass of KRb$_2$
($210$\unit{amu}). Within the tuning range of our Ti:Sapph-laser
(730\,$\unit{nm}$-1050\,$\unit{nm}$), WP oscillations are visible in
the pump-probe transients only between about 820\,\unit{nm}
(12200\wn{}) and 880\,\unit{nm} (11360\wn{}). As seen in the
PI-spectrum of Fig.~\ref{fig:FTKRb2}, there is an additional feature
at 12100\wn\, which does not coincide with the first excited quartet
states. We assume that this is a REMPI resonance of the pump pulse
since we observe the richest WP dynamics in this range. This
observation points to the existence of another quartet state in the
vicinity of 24200\wn.  We expect one frequency for each of the 3
vibrational modes of the $1^{4}$B$_{2}$ lowest quartet state as well
as for the two accessible excited states $3^{4}$A$_{1}$ and
$4^{4}$B$_{2}$.  As can be seen in Figure~\ref{fig:qx-all}c the
$3^{4}$A$_{1}$ and $4^{4}$B$_{2}$ excited states of KRb$_2$ feature a
conical intersection in the Q$_x$ coordinate at -0.2
\AA\,(11600$\wn$), which becomes an avoided crossing after inclusion
of SO coupling.  JT theory does not apply here since electronic
degeneracies do not occur at equilateral geometry and the vibrational
degeneracy is lifted.  Since the experimental excitation at 12050$\wn$
is well above the crossing point (and 450~cm$^{-1}$ above the minimum
of the 3$^4$A$_1$ state), the vibrational eigenstates of the
SO-corrected adiabatic potential curves used for the interpretation
differ significantly from their non-relativistic counterparts. The new
states, which can be labeled as E$_{1/2}$ states in the C$_{2v}$ spin
double group, are further referred to as the upper and lower excited
quartet PES.\newline The corresponding power spectrum
(Fig.~\ref{fig:FTKRb2}) is dominated by two lines at 23\wn{} and
31.5$\wn{}$. They are assigned to the Q$_x$ (26\wn) and Q$_y$ (31\wn)
modes of the $1^{4}$B$_{2}$ lowest quartet state and have long
lifetimes in the range 150\,ps as retrieved from the spectrogram. As
in the case of Rb$_3$ line shapes and spectrograms suggest
superimposed fast and slow decays.

The observed vibrational beats at 54.8$\wn$ and 56.6$\wn$ roughly
coincide with the calculated Q$_s$ ($52\wn$) and Q$_x$ (52$\wn$) modes
of the upper quartet PES. The Q$_s$ modes, however, have low
intensities and are shortlived for the examined alkali trimers. Since
the upper excited PES is also anharmonic, both observed frequencies
could originate from the same Q$_x$ mode.  As can be seen in
Figure~\ref{fig:qx-all}, we should preferentially excite higher
$v$-states with non-vanishing FC-factors from the lowest $v$-state in
the $1^{4}$B$_{2}$ potential. The calculated beat frequencies are
$\nu_{(1,2)}$=\,55\wn{}, $\nu_{(2,3)}$\,=54\wn{} ,
$\nu_{(3,4)}$\,=53\wn{} and $\nu_{(4,5)}=$\,54\wn{}.  The beat
frequency at 62\wn{} fits well to the Q$_y$ mode (62\wn{}) of the
upper excited PES.  Similar to our approach for Rb$_3$ we extract the
initial phases of the WP oscillations from the pump-probe transients
(see Table~\ref{tab:frequRb2K}). Within the errors, the states of the
$1^{4}$B$_{2}$ states show phases consistent with zero, whereas beats
associated with vibrations on the upper SO-coupled PES show an initial
phase of roughly $\pi$.  \newline In an alternative approach, we can
try to relate many observed frequencies to overtones of the Q$_x$ and
Q$_y$ modes of the $1^{4}$B$_{2}$ PES, which may be observable because
of the high line intensities.  We therefore assume the observed beat
at 46.1\wn{} to be a $\Delta v=2$ overtone of the Q$_x$ mode. The
initial phase of this component (0.6(2)\,\unit{2 \, \pi}), however,
does not match the Q$_x$ phase (0.05(9)\,\unit{2 \, \pi}) and is
consistent with the phases of the states on the upper excited
PES. Thus, this component may also be associated with Q$_s$, Q$_x$ or
Q$_y$ modes of the upper PES or with vibrations in even higher
states. The line at 62\wn{} can then alternatively be assigned to a
$\Delta v=2$ overtone of the $1^{4}$B$_{2}$ Q$_y$ mode. The remaining
line at 93\wn{} can be related to a $\Delta v=3$ overtone of the same
vibrational mode.  It is clear from Tabs.~2 and 3 in Paper~I that
FC-factors are non-zero only for the first few $v$-states in the
excited quartet potentials. Since the excitation is hundreds of
wavenumbers above the 3$^4$A$_1$ state minimum, we have to excite
$v>$10 in the lower excited PES which is very unlikely. We therefore
do not expect to observe any lines originating from the lower PES.

\begin{table}[ht]
\small
\begin{center}
  \caption{Observed vibrational beats of KRb$_2$ and calculated
    values.}
  \begin{tabular}{p{0.9cm}p{0.9cm}p{0.7cm}p{0.9cm}p{1.7cm}p{0.7cm}}\hline
		$\nu_{exp}$     & $\tau_1$    & $\tau_2$       & $\Phi$         & State		               &  $\nu_{theo}$  \\
    $[\wn]$         &$[\unit{ps}]$&$[\unit{ps}]$   &$[\unit{2 \, \pi}]$  & (mode)                 &  $[\wn]$\\
		\hline
					
			 -          		 & -        & -      	     & -         				 & $1^{4}$B$_{2}$(Q$_s$)        & 28     \\
			 \textbf{23.1(2)}         & 20(5)    & 136(44)      & 0.05(9)	       	 & $1^{4}$B$_{2}$(Q$_x$)        & 26     \\
			 \textbf{31.4(2)}  	     & 19(2)    & 155(30)      & 0.2(2) 		       & $1^{4}$B$_{2}$(Q$_y$)        & 31     \\
			 46.1(2)         & -        & 470(170)     & 0.6(3) 		       & $1^{4}$B$_{2}$(Q$_x)$        & 2$\times$26   \\
			 -               & -        & -            & -                 & $4^{4}$B$_{2}$(Q$_s$)        & 52     \\
			 54.8(5)     		 & -        & 82(12  	     & 0.6(2)    				 & $4^{4}$B$_{2}$(Q$_x)\,(3,4)$ & 52     \\
			 56.6(2)     		 & -        & 116(15)	     & 0.6(1)    				 & $4^{4}$B$_{2}$(Q$_x)\,(4,5)$ & 54     \\
			 62.0(3)     		 & 16(1)    & 240(110)     & 0.83(17)  				 & $4^{4}$B$_{2}$(Q$_y$)        & 62     \\
			 66.5		 				 & -				& -						 & - 								 & $1^{4}$B$_{2}$(Q$_x$)	      & 3$\times$26   \\
			 93   				   & - 				& - 			     & -					     	 & $1^{4}$B$_{2}$(Q$_y$)        & 3$\times$31   \\
			 \hline
			\end{tabular}		
		\label{tab:frequRb2K}
              \end{center}
\end{table}

\subsection{The K$_2$Rb trimer}
\label{k2rb}
\begin{figure}[ht]
  \begin{center}
     \includegraphics[width=0.45\textwidth]{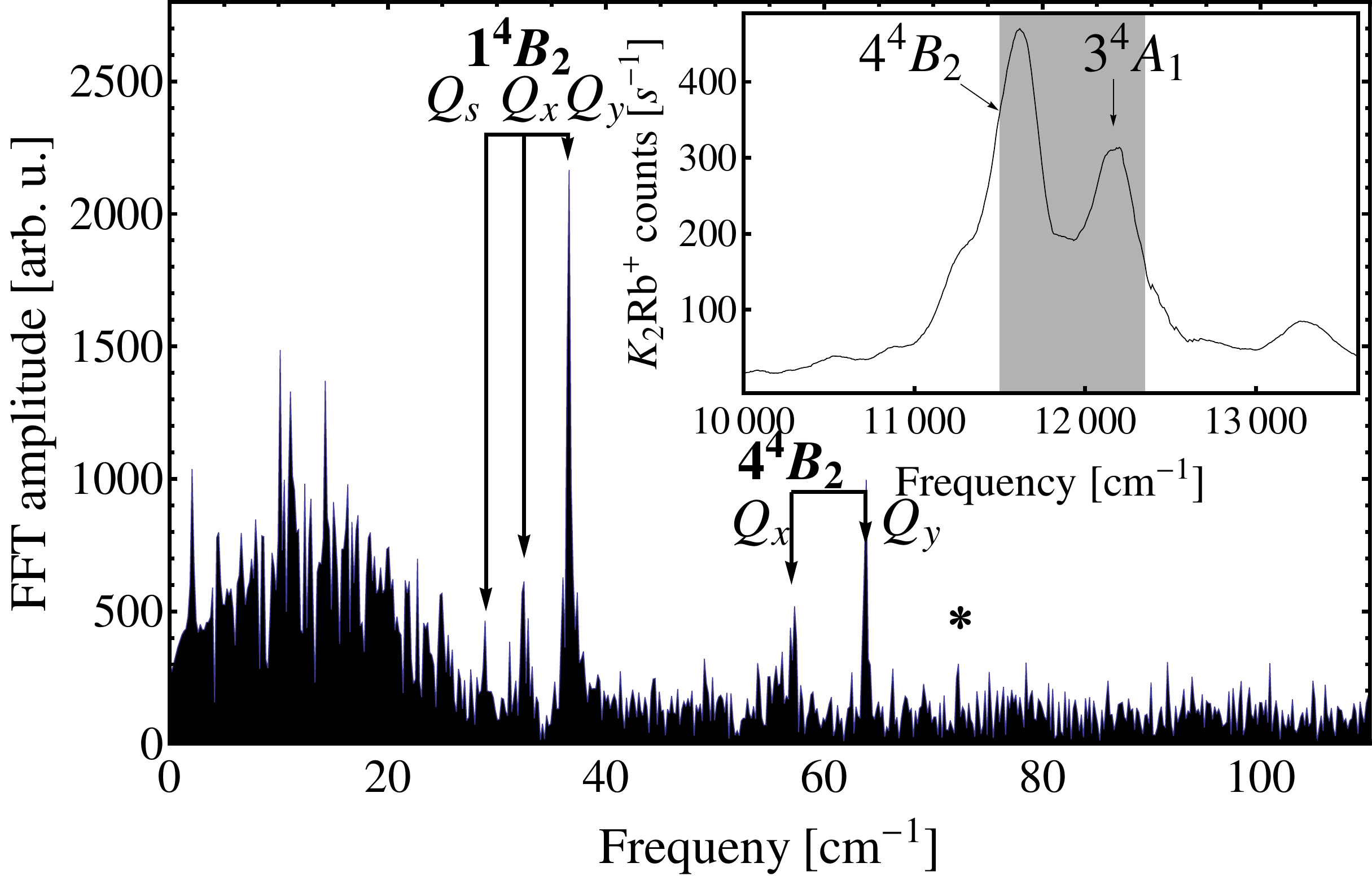}
     \caption{Fourier transform of the K$_2$Rb pump-probe
       photo\-ionization transient at 860\unit{nm}\,(11630\wn{}) with
       the corresponding photoionization spectrum. \label{fig:FTK2Rb}}
\end{center}
\end{figure}

Fourier spectra in the range from 810\,\unit{nm} (12346\,$\wn$) to
870\,\unit{nm} (11494\,\wn{}) were recorded for the lighter mixed
trimer K$_2$Rb (163.7\unit{amu}) which show clearly visible WP
dynamics around 860\,nm (11630\wn{}) as can be seen in
Fig.~\ref{fig:FTK2Rb}. The three lines at lower frequencies are
assigned to vibrations in the $1^{4}$B$_{2}$ lowest quartet state
where the component at 32.5\wn{} agrees reasonably well both with the
calculated Q$_s$ mode (34\wn{}) and the Q$_x$ mode (34\wn{}). Since
the symmetric stretch modes have been observed to be much weaker than
the asymmetric stretch and bending modes for all examined trimers we
favor the assignment to the Q$_x$ mode. Another weak line is found at
28.2\wn{}, but the assignment to the Q$_s$ mode is rather
speculative. Because of the high intensities of the femtosecond laser
in the range of 1\unit{GW/cm^2}, one could imagine multiphoton Raman
excitation, leading to excitation of higher $v$-states in the lowest
quartet potential. The latter proves to be quite anharmonic, and an
excitation to $v$=4-6 roughly matches the calculated frequencies
($\nu_{(4,5)}$\,=30.2\wn{}, $\nu_{(5,6)}$=\,29.5\wn{} and
$\nu_{(6,7)}$=\,28.8\wn{}).  The calculations yield a Q$_y$ frequency
of 35\wn{} which is in good agreement with the observed line at
36.6\wn{}. The beat note at 56.3\wn{} is quite close to the very
harmonic Q$_x$ (51\wn) mode of the $4^{4}$B$_{2}$-state. Finally, the
observed frequency at 63.8\wn{} is assigned to vibrations in the Q$_y$
mode (68\wn{}) of the $4^{4}$B$_{2}$-state. In this case, the
difference of several wavenumbers between the observed and calculated
frequency can be explained by the failure of the harmonic
approximation. The finite differences calculation of the first beat
frequencies are $\nu_{(0,1)}$=\,69\wn, $\nu_{(1,2)}$=\,67.7\wn{} and
$\nu_{(2,3)}$=\,66.5\wn{} and $\nu_{(3,4)}$=\,61.6\wn{}. We therefore
find a best match for beat frequencies $\nu_{(2,3)}$ or
$\nu_{(3,4)}$. The Q$_s$ frequency calculation yields a value of
48\wn{} and the spectrum suggests that there is a very weak line at
this position (see Fig.~\ref{fig:FTK2Rb}). However, the noise is too
large to make a definitive assignment.\newline As for the case of
KRb$_2$, we can try to alternatively relate some of the observed
frequencies to $\Delta v=2$-overtones of the modes in the lowest
quartet PES $1^{4}$B$_{2}$.  Following these lines, the observed
frequency difference at 56.3\wn{} could be the overtone of the Q$_s$
mode (28\wn{}), and likewise the beat frequency at 63.8\wn{} could be
the Q$_x$ (32.5\wn{}) overtone.  The very weak frequency component at
72.5\wn{} might then be the corresponding $\Delta v=2$-beat of the
Q$_y$ mode (36.6\wn{}). Unlike for the case of KRb$_2$, though, the
intensities of the presumed overtones are mostly higher here than the
fundamental frequencies which is not easily understood.

Since the excitation laser frequency of 11628\wn{} is tuned to the
transition $4^{4}$B$_{2}\leftarrow1^{4}$B$_{2}$ we do not expect
vibrations of the $3^{4}$A$_{1}$-state which is roughly 150\wn{}
higher in energy and out of range for the given excitation
bandwidth. As for the other species, the WP phases were extracted from
the spectrogram data yielding about 0.3\,\unit{2 \, \pi} for the
$1^{4}$B$_{2}$ potential. However, the beat at 36.0\wn, which is
assigned to the Q$_y$ mode, is not consistent with this value
(0.18(2)\,\unit{2 \, \pi}). The $4^{4}$B$_{2}$ excited quartet-state
beats are found to have phases of about 0.6\,\unit{2 \, \pi}.

\begin{table}[ht]
\small
\begin{center}
\caption{Observed vibrational beats of K$_2$Rb and calculated values.}
  \begin{tabular}{p{0.9cm}p{0.9cm}p{0.9cm}p{0.9cm}p{1.3cm}p{0.9cm}}\hline
		   $\nu_{exp}$     & $\tau_1$    & $\tau_2$       & $\Phi$         & State		            &  $\nu_{theo}$  \\
       $[\wn]$         &$[\unit{ps}]$&$[\unit{ps}]$   &$[\unit{2 \, \pi}]$  & (mode)               &  $[\wn]$\\
			 \hline
			 28.2 			     & - 	         & -	 				    & 0.30(2)    		 & $1^{4}$B$_{2}$(Q$_s$)  & 34    \\
			 32.5(5)      	 & 26(29) 		 & -	            & 0.35(2)		     & $1^{4}$B$_{2}$(Q$_x$)  & 34    \\
			 \textbf{36.6(2)}  	     & 28(5)       & 440(200)       & 0.18(2)   	   & $1^{4}$B$_{2}$(Q$_y$)  & 35    \\
			 -          	   & -   				 & -	            & -	    	       & $4^{4}$B$_{2}$(Q$_s$)  & 49    \\
			 \textbf{56.3(4)}         & -   				 & -	            & 0.7(2)	   	   & $4^{4}$B$_{2}$(Q$_x$)  & 51    \\
			 \textbf{63.8(2)}        & 38(2)	     	 &	-	   	        & 0.56(16)       & $4^{4}$B$_{2}$(Q$_y$)  & 68    \\
			 \hline
		\end{tabular}		
		\label{tab:frequRbK2}
\end{center}
\end{table}

\begin{figure}[ht]
  \begin{center}
    { \includegraphics[width=0.45\textwidth]{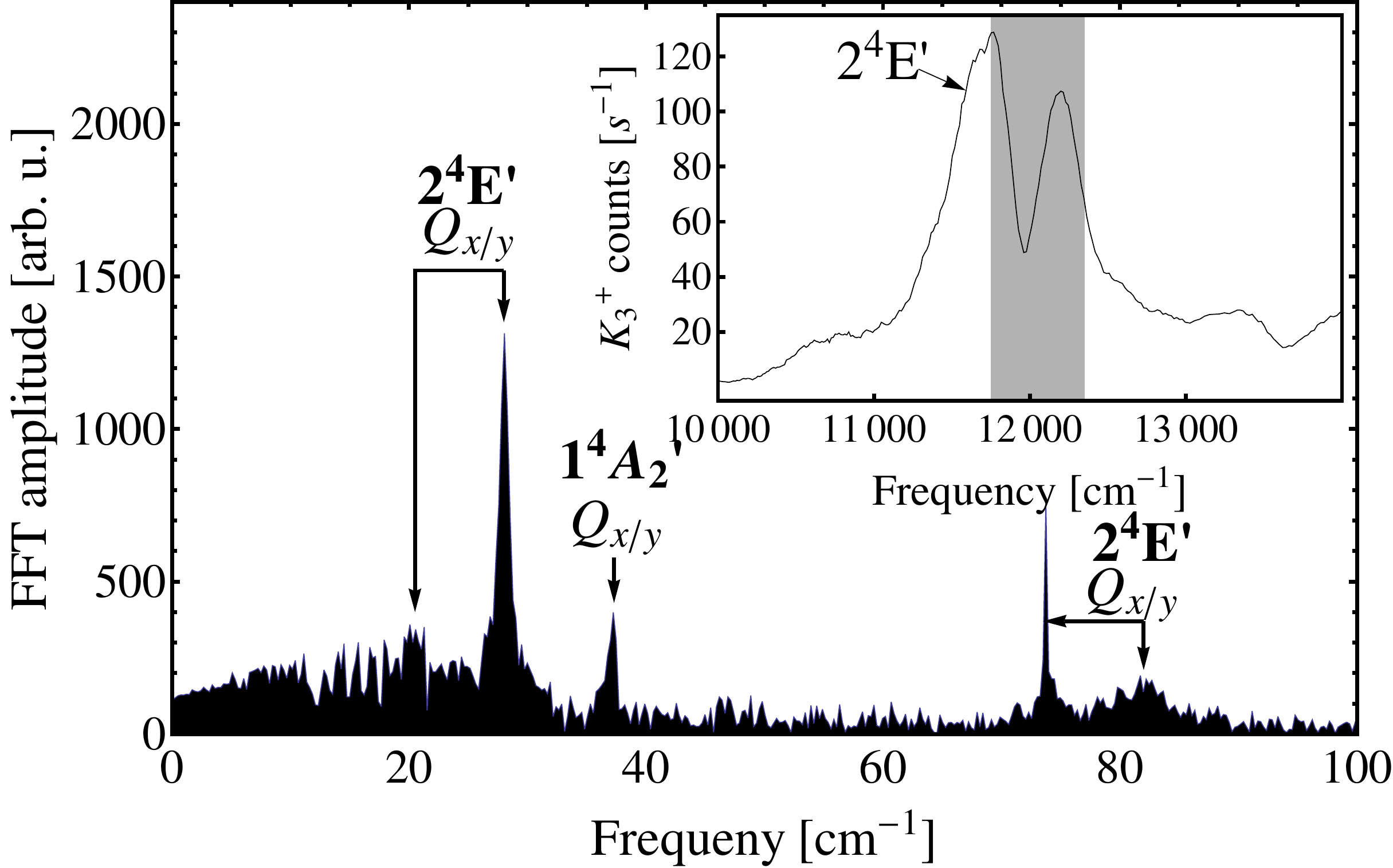}
      \caption{Fourier transform of the K$_3$ pump-probe
        photoionization transient at 840\,$\unit{nm}\,(11835\wn{})$. The range of
        the observed WP-dynamics is visualized in gray in the photoionization
        spectrum (inset) \label{fig:FTK3}}}
\end{center}
\end{figure}

\subsection{The K$_3$ trimer}
\label{k3}
The Fourier spectrum of the potassium trimer K$_3$ (117.3\unit{amu})
at 840~nm (11835\,$\wn$) depicted in Fig.~\ref{fig:FTK3} shows some
resemblance to its heavier cousin Rb$_3$. However, peak assignment
turns out to be more difficult here. Our suggestions are summarized in
Table~\ref{tab:frequK3}.

\begin{table}[ht]
\small
\begin{center}
\caption{Observed vibrational beats of K$_3$ and calculated values.}
\begin{tabular}{p{0.8cm}p{0.5cm}p{0.8cm}p{0.7cm}p{2cm}p{0.7cm}}\hline
		   $\nu_{exp}$    & $\tau_1$     & $\tau_2$       & $\Phi$          & State		                                 &  $\nu_{theo}$  \\
       $[\wn]$        &$[\unit{ps}]$ &$[\unit{ps}]$   &$[\unit{2 \, \pi}]$   & (mode)                                   &  $[\wn]$\\ 						 \hline
       20     		    & -            & -              & -       	      &$2^{4}$E$^{\prime{}}(1,3)$                &  22    \\
			 \textbf{28(2)}        	& 18 			     & -   	          & 0.14(3)     	  &$2^{4}$E$^{\prime{}}(1,4)$ or             &	31       \\
			 -   						& - 					 & -   						& -           	  &$1^{4}$A$_{2}^{\prime}($Q$_{s})\,(3,4)$   &	33.8    \\
			 \textbf{37.1(2)}		    & 17(6)        & 58(25)         & 0.26(20)	      &$1^{4}$A$_{2}^{\prime}($Q$_{x/y})\,(3,4)$ &  36.1    \\
			 -            	& -            & -     		      & -          		  &$2^{4}$E$^{\prime{}}($Q$_{s})$            &  53      \\
			 \textbf{73.8(2)}        & 21(7)        & 180(120)       & 0.14(13) 	      &$2^{4}$E$^{\prime{}}(1,6)$                &  72      \\
			 81.7(2)        & 7(5)   			 & -              & 0.48(3)      	  &$2^{4}$E$^{\prime{}}(1,8)$	               &  81		    \\
			 110.0(5)       & 9(5)         & 42             &	-						    & -	                  	                   & -       \\ \hline
\end{tabular}		
\label{tab:frequK3}
\end{center}
\end{table}

Vibrational WP dynamics is observed in a window between 810\,nm
(12345\wn{}) and 850\,nm (11765\wn{}) which corresponds to the
$2^4$E$^{\prime{}}\leftarrow 1^4$A$_2^{\prime{}}$
transition,\cite{Reho:2001,hauser10jcp} but again the richest spectra
are obtained within a narrow interval of about 10\,nm around 840\,nm
(11905\wn{}). When comparing to the Rb$_3$ spectrum one is led to
assign the component at 28~\wn{} to Q$_{s}$ and the beat note at
37~\wn{} to the Q$_{x/y}$ mode of the lowest quartet state
$1^{4}$A$_2^{\prime{}}$. Beside the qualitative resemblance of the
spectra, the wavelength dependence and the lifetime of the observed
beats are similar to their counterparts in Rb$_3$. The tentatively
assigned Q$_{x/y}$ beat is visible over a wide range of excitation
wavelengths (400~\wn{}) whereas the peak at 28\wn{} only appears in a
window of some 100~\wn{} around 11900~\wn{}. The lifetime of the Q$_s$
beat is~16\,$\unit{ps}$ in the case of Rb$_3$ and~17\,$\unit{ps}$ for
the discussed beat component whereas the 37\wn{} beat is observed over
longer delay times (58(25)\,$\unit{ps}$).

\begin{table}[ht]
	\centering
	\caption{Calculated beat frequencies (in cm$^{-1}$) of the Q$_s$
          and Q$_{x/y}$-modes in the $1^{4}_{2}$A$_{2}'$ lowest quartet
          state of K$_3$.}
		\begin{tabular}{cccccc}									Mode     & $v=0,1$            &$v=1,2$            &$v=2,3$            &$v=3,4$                      \\
			\hline
			\hline   Q$_s$     & 37					        & 35.9				  	   & 34.8						   & 33.8   				   				  \\
			\hline 	 Q$_{x/y}$ & 34.1			  	      & 35.5					     & 35.9  					   & 36.1					     				  \\
		\end{tabular}		
		\label{tab:vibfrequ}
\end{table}

The calculated harmonic frequencies of the Q$_{x/y}$ (33\wn{}) and
Q$_s$ (37\wn{}), however, do not support this interpretation. A better
agreement is achieved when assuming that vibrational states with $v>3$
are populated in the anharmonic PES (see Tab.~\ref{tab:vibfrequ})
which is imaginable because of the high laser intensities that may
drive multiphoton transitions.  Alternatively, the beat at 28\wn{} can
be related to the lowest vibrational levels of the excited
$2^{4}$E$_{2}^{\prime}$ state (see Figure~\ref{fig:qx-all}a and
Tab.~\ref{tab:eig-k}). For the first few eigenstates, namely 1, 3, 4,
6 and 8 there is good FC overlap with the lowest vibrational
eigenfunction of the $1^{4}$A$_{2}^{\prime}$ state and there is a
reasonably good match with the calculated component (1,4)
(31~\wn{}). Although difficult to detect because of its low frequency
there is some experimental evidence for another small peak around
20~\wn{} which might be related to the beat (1,3)
(22~\wn{}). According to Ref.\citenum{hauser10jcp} these two smaller
beats are caused by coherent excitation of the lowest vibrational
states of different spin surfaces, whereas the difference frequencies
(1,6) (72~\wn{}) and (1,8) (81~\wn{}) involve fundamental
vibrations. The two latter beat components are in good agreement with
the experimental peaks at 73.8~\wn{} and 81.7~\wn{}. As for the case
of Rb$_3$, it has to be mentioned that the observed beat at 73.8~\wn{}
can also be related to an overtone of the Q$_{x/y}$ in the lowest
quartet PES $1^{4}$A$_{2}^{\prime}$.  The component is observed over a
larger wavelength range (400\wn{}) than the observed width of the cw
excitation (200\wn{}), pointing at a RIRS transition.  Just like for
Rb$_3$, the laser bandwidth and intensity could explain this
observation.

As expected, the theoretical value of 67\wn{} for the Q$_{x/y}$-mode
in the JT distorted $2^{4}$E$^{\prime{}}$ state does not occur due to
SO coupling. The harmonic frequency (52\wn{}) for Q$_s$ in the
$2^{4}$E$^{\prime{}}$ state does not occur either.

\begin{table}[ht]
\small
\begin{center}
  \caption{Positions of vibronic levels (in \wn{}) in the
    JT-distorted 2$^4$E$^{\prime}$ electronically excited state of
    K$_3$. The harmonic frequency of the degenerate Q$_{x/y}$ mode is
    67\wn{}.     \label{tab:eig-k}}
\begin{tabular*}{0.45\textwidth}{@{\extracolsep{\fill}}cccc} \hline
  Nr.&abs. position    &rel. position$^a$& intensity \\ \hline
  \textbf{1}   &      \textbf{11903}   &      \textbf{48}    &      \textbf{3.97}   \\
  2   &      11914   &      59    &      3.59   \\
  \textbf{3}   &      \textbf{11925}   &      \textbf{70}    &      \textbf{3.50}   \\
  \textbf{4}   &      \textbf{11934}   &      \textbf{79}    &      \textbf{3.55}   \\
  5   &      11969   &     114    &      0.00   \\
  \textbf{6}   &      \textbf{11975}   &     \textbf{120}    &      \textbf{0.50}   \\
  7   &      11979   &     124    &      0.00   \\
  \textbf{8}   &      \textbf{11984}   &     \textbf{129}    &      \textbf{0.21}   \\
  9   &      11989   &     134    &      0.00   \\
  1   &      11994   &     139    &      0.13   \\
  11  &      11997   &     142    &      0.00   \\
  12  &      12003   &     148    &      0.13   \\
  13  &      12034   &     179    &      0.00   \\
  14  &      12041   &     186    &      0.39   \\
  15  &      12044   &     189    &      0.00   \\
  16  &      12046   &     191    &      0.00   \\
  17  &      12049   &     194    &      0.38   \\
  18  &      12053   &     198    &      0.00   \\
  19  &      12054   &     199    &      0.00   \\
  20  &      12058   &     203    &      0.37   \\
  21  &      12061   &     206    &      0.00   \\
  22  &      12063   &     208    &      0.00   \\
  23  &      12066   &     211    &      0.37   \\ \hline
 \end{tabular*}		
\end{center}
$^a$Energetic difference to the point of the conical intersection,
assuming no zero-point energy in Q$_s$.
\end{table}

\subsection{Conclusion}
\label{sec:conclusion}
In the presented combined experimental and theoretical study we
compare measured and \emph{ab initio} calculated vibrational
frequencies of quartet states of all triatomic molecules composed of K
and Rb atoms. These molecules are formed on the surface of helium
nanodroplets and interrogated using femtosecond Fourier
spectroscopy. This technique circumvents the considerable spectral
perturbations induced by the helium environment that limit the
resolution of linear spectroscopy by probing vibrational coherences in
the time-domain. Thus, vibrational beat frequencies can be determined
with high precision of the order 0.2\wn{}.\newline In contrast to the
straight forward assignment of pump-probe power spectra recorded for
diatomic alkali molecules\cite{Claas:2006}\cite{Mudrich:2009}, the
interpretation of vibrational spectra of alkali trimers turns out to
be much more involved due to a complex interplay of JT and
SO-couplings. Moreover, the amplitudes of the observed beat components
cannot be analyzed easily because of the Raman-type multiphoton
excitations and ionizations via unknown higher electronic
states. Therefore, in many cases the assignment of experimental
frequencies to calculated values has to be considered as tentative.
\newline Nevertheless, we suggest assgnments of the measured beat
frequencies to vibrational modes of the lowest and excited quartet
states by taking several experimental observables into consideration,
including line positions and widths, laser intensity-dependences and
mode-specific ultrafast dynamics. The agreement with vibrational
frequencies obtained from high-level \emph{ab initio} calculations
that include Jahn-Teller and spin-orbit perturbations is satisfactory
in the cases of Rb$_3$, KRb$_2$, and K$_2$Rb. Most prominent lines in
the Fourier spectra of these molecules are the asymmetric stretch and
bending modes Q$_{x/y}$ of the lowest quartet state excited by
impulsive Raman scattering, followed by the Q$_{x/y}$-modes of the
accessible excited electronic states. The breathing mode Q$_s$ is only
visible in the spectrum of Rb$_3$, where it is significantly broadened
due to fast dephasing within a few ps. Intramolecular vibrational
relaxation, intersystem-crossing or vibrational relaxation by coupling
to the helium environment may be at the origin of this fast
dynamics. In contrast, the Q$_{x/y}$-modes feature long coherence
times of the order of 100\,\unit{ps}. In the case of K$_3$ several
frequencies can not be assigned unambiguously. The presented spectroscopic data are expected to serve
as valuable input information for upcoming experiments aiming at
forming ultracold triatomic molecules in optical traps and lattices.

\section*{Acknowledgments}
We thank C. Callegari for valuable discussion and DFG for financial
support. AWH gratefully acknowledges support from the Graz Advanced
School of Science, a cooperation project between TU Graz and the
University of Graz, and the Austrian Science Fund (FWF, Grant
No. P19759-N20).

\providecommand*{\mcitethebibliography}{\thebibliography}
\csname @ifundefined\endcsname{endmcitethebibliography}
{\let\endmcitethebibliography\endthebibliography}{}

\end{document}